\documentclass[a4paper,fleqn]{cas-sc}

\usepackage[authoryear]{natbib}

\usepackage{lineno}

\usepackage{xcolor}
\usepackage[acronym]{glossaries}
\usepackage{listings}
\usepackage{subfig}

\newacronym{qc}{QC}{Quality Control}
\newacronym{saqc}{SaQC}{System for automated Quality Control}

\def\tsc#1{\csdef{#1}{\textsc{\lowercase{#1}}\xspace}}
\tsc{WGM}
\tsc{QE}

\begin{document}
\let\WriteBookmarks\relax
\def\floatpagepagefraction{1}
\def\textpagefraction{.001}

\shorttitle{NORA-Surge hindcast}    

\shortauthors{{NMK, PT, JR, OJA, {\O}S, \O}B}  

\title [mode = title]{NORA-Surge: A storm surge hindcast for the Norwegian Sea, the North Sea and the Barents Sea}   



%

\author[1]{Nils Melsom Kristensen}[orcid=0000-0002-2494-6509]
\credit{Conceptualization, Data curation, Formal analysis, Investigation, Validation, Visualization, Writing – original draft, Writing – review \& editing}
\ead{nilsmk@met.no}
\author[1,3]{Paulina Tedesco}[]
\credit{Conceptualization, Data curation, Formal analysis, Investigation, Validation, Visualization, Writing – original draft, Writing – review \& editing}
\author[1]{Jean Rabault}[]
\credit{Conceptualization, Data curation, Formal analysis, Investigation, Software, Validation, Writing – original draft, Writing – review \& editing}
\author[4]{Ole Johan Aarnes}[]
\credit{Conceptualization, Formal analysis, Writing – original draft, Writing – review \& editing}
\author[1]{{\O}yvind Saetra}[]
\credit{Conceptualization, Supervision, Writing – review \& editing}
\author[1,2]{{\O}yvind Breivik}[]
\credit{Conceptualization, Funding acquisition, Project administration, Supervision, Writing – original draft, Writing – review \& editing}
%

\cormark[1]
\cortext[1]{Corresponding author}

\affiliation[1]{organization={Norwegian Meteorological Institute},
            addressline={P.O. Box 43 Blindern}, 
            city={Oslo},
            postcode={NO-0313}, 
            country={Norway}}

\affiliation[2]{organization={University of Bergen},
            city={Bergen},
            country={Norway}}

\affiliation[3]{organization={University of Oslo},
            city={Oslo},
            country={Norway}}  
            
\affiliation[4]{organization={Equinor ASA},
            city={Bergen},
            country={Norway}}


\begin{abstract}
Knowledge about statistics for water level variations along the coast due to storm surge is important for the utilization of the coastal zone. An open and freely available storm surge hindcast archive covering the coast of Norway and adjacent sea areas spanning the time period 1979-2022 is presented. The storm surge model is forced by wind stress and mean sea level pressure taken from the non-hydrostatic NORA3 atmospheric hindcast. A dataset consisting of observations of water level from more than 90 water level gauges along the coasts of the North Sea and the Norwegian Sea is compiled and quality controlled, and used to assess the performance of the hindcast. The observational dataset is distributed in both time and space, and when considering all the available quality controlled data, the comparison with modelled water levels yield a mean absolute error (MAE) of $9.7$~cm and a root mean square error (RMSE) of $12.4$~cm. Values for MAE and RMSE scaled by the standard deviation of the observed storm surge for each station are $0.42$ and $0.54$ standard deviations, repsectively. When considering the geographical differences in characteristics of storm surge for different countries/regions, the values of MAE and RMSE are in the range $5.7-13.9$~cm and $7.6-17.8$~cm respectively, and $0.33-0.46$ and $0.42-0.59$ standard deviations for the scaled values.  The minimum and maximum values for water level in the hindcast are $-2.60$~m and $3.92$~m. In addition, 100-year return level estimates are calculated from the hindcast, with minimum and maximum values of, respectively, $-2.75$~m and $3.98$~m. All minimum and maximum values are found in the southern North Sea area.
\end{abstract}


\maketitle

\section{Introduction}\label{sec:introduction}
Storm surges are driven by the passing of synoptic weather systems. A simple rule of thumb states that a reduction of 1 hPa in surface air pressure leads to an increase of 1 cm in water level from the inverse barometric effect alone. In addition, frictional forces in combination with the Coriolis effect will lead to Ekman transport which can cause convergence and divergence of water masses along coastlines. More generally, such perturbations of the sea surface will tend to manifest themselves as Kelvin waves, travelling with the coast to their right in the northern hemisphere. As these waves have spatial and  temporal scales comparable to the astronomical tides, they have historically been known as tidal surges, but their generating mechanism is of course very different. More importantly, storm surges are only as predictable as the synoptic weather phenomena that cause them. There exists a number of well-documented water level forecast systems in Europe and elsewhere (e.g. \citealt{glahn2009role, zijl:etal:2013, kristensen22}). The simplest represent the ocean as a two-dimensional barotropic field with a free surface. This is also how we model the water level in this study. As Kelvin waves propagate as shallow-water waves whose phase and group speed are $\sqrt{gH}$, where $H$ is the local depth and $g \approx 9.81\,\mathrm{m\,s^{-2}}$ the earth's gravitational acceleration, the amplitude and propagation speed of storm surges is sensitive to errors in both the gradient and the mean level of the bathymetry.

The availability of storm surge hindcasts for the Norwegian Sea, the North Sea and the Barents Sea region with sufficient spatial resolution, and spatiotemporal coverage is scarce, as shown by \cite{FERNANDEZMONTBLANC2020105367}. 
Long hindcast integrations are necessary to assess water level extremes in coastal regions. To do this realistically requires high-quality and high-resolution atmospheric forcing and sufficiently high spatial resolution for the storm surge model. 
Storm surge hindcast archives serve mainly three practical applications. The first is to properly account for the overall statistical distribution of the water level, and the other is to assess the extreme value distribution, i.e., the return values for extreme high and low water level. In addition to these statistical measures, the scarcity of observational records means that accurate reconstructions of past events are of societal value in themselves. High-quality hindcast archives are also essential for establishing a baseline against which we can estimate the future projections of changes to the storm surge climate \citep{bernier24}. The objective of this article is to present such a open and freely available historical hindcast and to assess its validity for extreme value estimation of water level in the North Sea, the Norwegian Sea and the Barents Sea.

This article is structured as follows. In Section \ref{sec:obs_dataset} we present the observational dataset used in the assessment of the hindcast. In Section \ref{sec:hindcast_setup} we describe the storm surge hindcast model setup and the the atmospheric forcing used. In Section \ref{sec:hindcast_eval_stats} the performance is evaluated against in-situ observations and we present the extreme value estimates and hindcast statistics for storm surge. A summary and some concluding remarks are found in Section \ref{sec:discussion_and_conclusion}.

\section{Observational dataset}\label{sec:obs_dataset}

\begin{figure}[ht]
    \centering
    \includegraphics[width=0.5\linewidth]{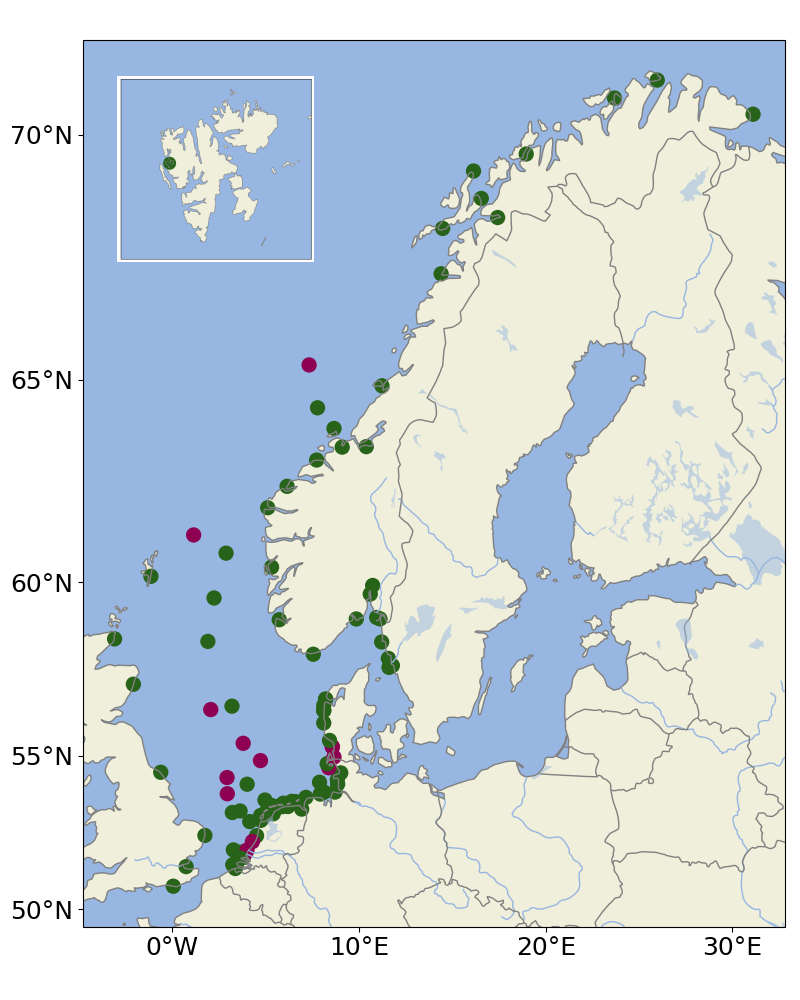}
    \caption{Observation locations. Green dots indicate stations used in the analysis, and red dots indicate stations where data are available but that were discarded due to quality issues.}
    \label{fig:obs_map}
\end{figure}

Water level gauges from the coast of Norway (stations with NO prefix, also used in the study by \citealt{tedesco23}), Norwegian oil platforms (WL), Denmark (DK), Sweden (SW), Germany (DE), the Netherlands (NL), and the United Kingdom (UK) make up our observational data. Unfortunately, the license agreements do not allow open redistribution of the data we gathered. However, we share the code used for parsing and assembling the data from these sources into a single dataset (see Appendix A). We also summarize the source of the data from each country in  Appendix B. 

The data provided come from a variety of water level gauges. The location of the gauges and the associated measurement stations are shown in Figure \ref{fig:obs_map}. The time spans of the time series vary from station to station (see Figure \ref{fig:time_cover_diff_obs_mod}). Some stations have continuous recordings going back to 1915 (for some of the UK stations), though our dataset is built starting from 1979 onwards. As a consequence of the large time span covered, the kind of gauge, temporal resolution, and water level resolution of the measurements is not homogeneous in time even sometimes for a single individual station. Moreover, some data sources provide a quality flag as part of the data downloaded (depending on the data source and station, this may be either an automatic quality flag, or a human determined expert quality flag), while others only provide raw data without quality flag. We have, therefore, performed a detailed quality control cycle on all the data, which includes quality checking of individual measurements through both flag checking and statistical analysis, interpolation on a common time base, linear detrending, visual inspection, automated outlier detection, and a series of quality checks on running windows of observational data. The quality control removes approximately $9\%$ of the total observation data. The technical details of how this was performed are presented in Appendix B.

A tidal analysis was performed with the CNES pangeo-pytide python package \footnote{\url{https://github.com/CNES/pangeo-pytide}} in order to estimate the tidal signal at each station and to detide the observed water level. For this, the time series obtained for each station is used to obtain the modal coefficients for the default 67 tidal modes used by pytide \footnote{the corresponding modes are: O1, P1, K1, 2N2, Mu2, N2, Nu2, M2, L2, T2, S2, K2, M4, S1, Q1, Mm, Mf, Mtm, Msqm, Eps2, Lambda2, Eta2, 2Q1, Sigma1, Rho1, M11, M12, Chi1, Pi1, Phi1, Theta1, J1, OO1, M3, M6, MN4, MS4, N4, R2, R4, S4, MNS2, M13, MK4, SN4, SK4, 2MN6, 2MS6, 2MK6, MSN6, 2SM6, MSK6, MP1, 2SM2, Psi1, 2MS2, MKS2, 2MN2, MSN2, MO3, 2MK3, MK3, S6, M8, MSf, Ssa, Sa}, and to perform tidal prediction on the same time base as the observations. By subtracting the harmonic tide analysis estimates obtained with pytide from the in-situ observations, we compute the (non-tidal) weather-induced residuals that are used to evaluate the storm surge model quality. This is further illustrated in Appendix B.

Following these steps, we have constituted a dataset of quality controlled sea level observations that are available on a common time base, together with tide elevation data derived from modal analysis. These data are used in the following as a ground truth estimate for validation of the numerical model.

\section{NORA-Surge: Hindcast setup}\label{sec:hindcast_setup}
The hindcast integration covers the period 1979 to 2022, inclusively. The model domain covers the North Sea, the Norwegian Sea and the Barents Sea (see Figure~\ref{fig:domain_nordic4}). The spatial resolution is approximately 4 km on a polar stereographic projection. Fields of storm surge are archived hourly. No data assimilation has been applied, and the hindcast has been run as one continuous run with yearly restarts from model restart files.

\subsection{The ocean model}\label{sec:model}
\begin{figure}[ht]
    \centering
    \includegraphics[width=0.85\linewidth]{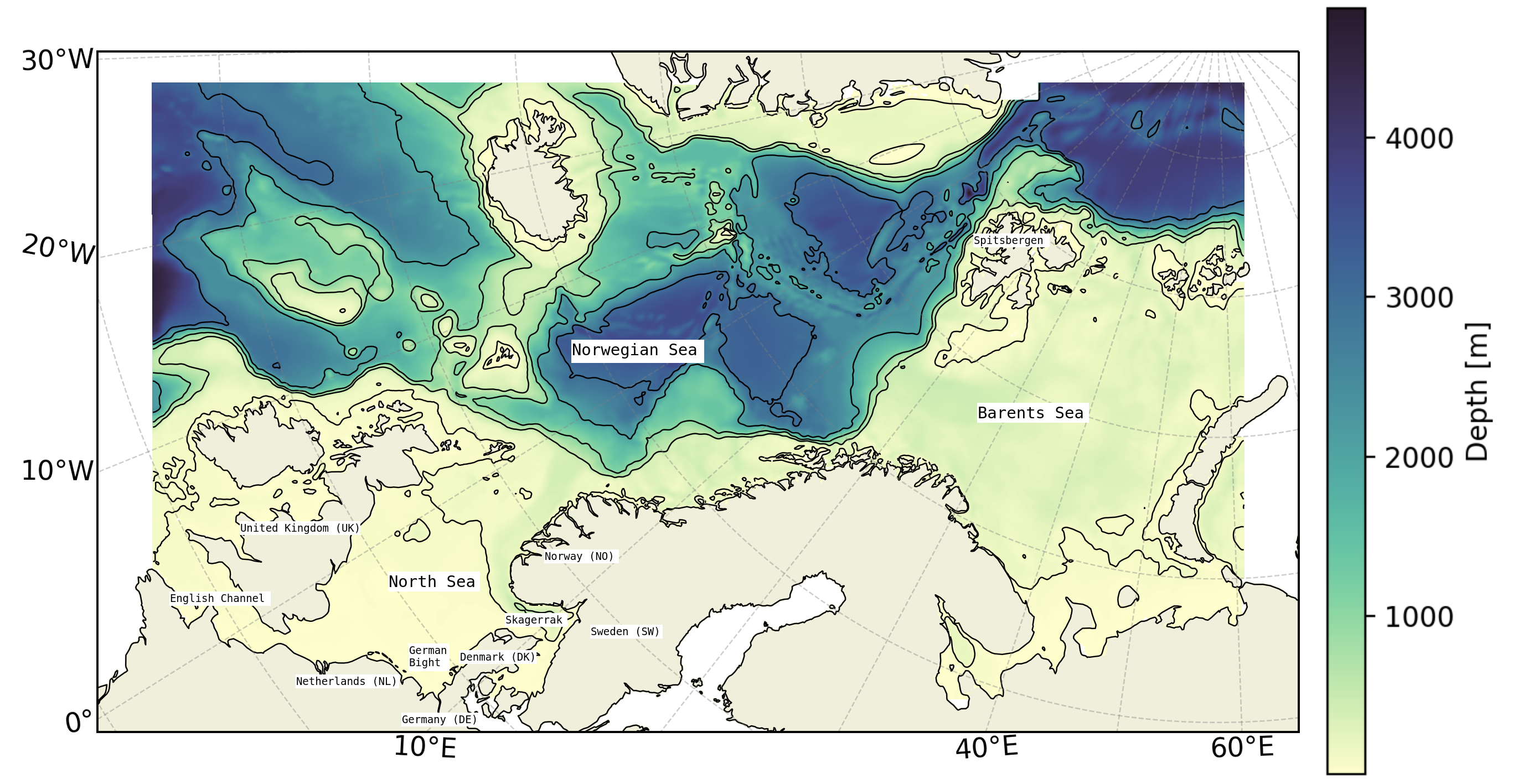}
    \caption{The ROMS Nordic-4km model domain covering the North Sea, the Norwegian Sea and the Barents Sea. Color scale and contours indicate model bathymetry.}
    \label{fig:domain_nordic4}
\end{figure}

The Regional Ocean Model System (ROMS, see \citealt{shchepetkin05}) is used for operational storm surge forecasting for Norwegian waters, as described by \citet{kristensen22}. The model is set up to run in two-dimensional barotropic mode and the model domain is depicted in Figure \ref{fig:domain_nordic4}. The governing model equations are thus the shallow water equations as described in \citet{haidv:etal:2008}. Let the total height be $h = H + \zeta$, where the $H$ is the equilibrium depth (Mean sea level is chosen as reference level in our simulations) and $\zeta$ is the sea surface deviation from the mean. By averaging the velocity over the total height $h$ we get \citep{kristensen22,tedesco23},

\begin{equation}
    \frac{\partial \mathbf{U} }{\partial t}  + \nabla_\mathrm{H} \cdot (\frac{ \mathbf{UU}}{h}) + f  \mathbf{k} \times  \mathbf{U} = -gh \nabla \zeta + \rho_0 ^{-1} (\boldsymbol{\tau}_\mathrm{s} - \boldsymbol{\tau}_\mathrm{b}) + \mathbf{X}.
    \label{eq:swe_1}
\end{equation}

Here  $f$ is the Coriolis parameter, $\rho_0$ is the sea water density which is kept constant, $\boldsymbol{\tau}_\mathrm{s}$  and  $\boldsymbol{\tau}_\mathrm{b}$ are the surface (wind) and bottom stress, respectively, and $\mathbf{X}$ is the horizontal diffusion and internal mixing. The subscript H denotes horizontal differentiation.
The evolution of the water level can be written

\begin{equation}
    \frac{\partial h}{\partial t} + \nabla_\mathrm{H} \cdot \mathbf{U}= 0.
    \label{eq:swe_2}
\end{equation}

No astronomical tides or boundary values from an external model are included in the runs. The decision to exclude astronomical tides as input at the open boundaries is made based on the well known problem of simulating correct tidal phases and amplitudes in numerical models. As mentioned by \citet{kristensen22}, tides and storm surge are non-linearly dependant, and can not be completely separated. Hence, this choice could lead to reduced quality of the hindcast in shallow regions with large tidal amplitudes where the non-linear tide-surge interactions are important (see \citet{horsburgh:2007}). On the open boundaries, that are formulated using the Chapman condition for two-dimensional momentum \citep{chapman:1985} and the Flather condition \citep{flather:1976} for the free surface, we only impose the inverse barometric water level to balance the interior solution,

\begin{equation}
     \zeta^\mathrm{IB} = \frac{1}{\rho_0 g}\left(\overline{p}_\mathrm{a} - p_\mathrm{a}\right).
\end{equation}

Here, $p_\mathrm{a}$ is the sea level pressure and $\overline{p}_\mathrm{a} = 1013.25\,\mathrm{hPa}$ is the global average sea level pressure. See \citet{kristensen22} for further details on the model setup.

\subsection{Atmospheric forcing---the NORA3 atmospheric hindcast}\label{sec:forcing}
The NORA3 hindcast \citep{haakenstad21nora3,haakenstad22} is a non-hydrostatic high-resolution atmospheric hindcast covering the North Sea, the Norwegian Sea and the Barents Sea. It is convection resolving and has a horizontal resolution of approximately 3~km. The spatial coverage ensures almost total coverage of the entire ocean model domain for the NORA-Surge hindcast, except for a small snippet of the north-western part that falls outside the NORA3 domain. Here ERA5 \citep{hersbach20era5} is used. This ensures dynamically consistent atmospheric forcing fields since NORA3 uses ERA5 as its host analysis \citep{haakenstad21nora3,haakenstad22}.
The combined fields of mean sea level pressure (MSLP) and 10~m wind stress calculated using the Charnock relation \citep{charnock:1955} are used to force the ocean model. The NORA3 hindcast has been found to yield realistic wind fields over the open ocean \citep{bre22}. As the model is non-hydrostatic, its wind and pressure fields are naturally more variable than those of the host reanalysis, ERA5. 

\section{NORA-Surge: Hindcast evaluation and statistics}\label{sec:hindcast_eval_stats}
\subsection{Model performance evaluated against observations}\label{sec:obs_vs_model}
The hindcast dataset has been extensively evaluated and validated against the observational dataset obtained in Section \ref{sec:obs_dataset}. In all the following analyses, the mean sea level for the entire hindcast period for each grid point has been subtracted from the modeled water level in order to establish mean sea level as the reference level. In order to establish a measure for the storm surge observation, the tidal predictions are subtracted from the time series for total water level for each of the stations. As mentioned in Section \ref{sec:model} and by \cite{idier:2019} and \cite{horsburgh:2007}, the "pure weather effect" can not be completely separated from the total water level due to non-linear interactions between different factors contributing to the total water level at the coast. Hence, these non-linear effects are one of the possible sources of errors in all the following analyses. The work performed in the present study does not allow us to quantify the contribution to the total error from the different sources.

\begin{figure}[ht]
  \centering
  \subfloat[Standard deviation observations and hindcast]{\includegraphics[scale=0.5]{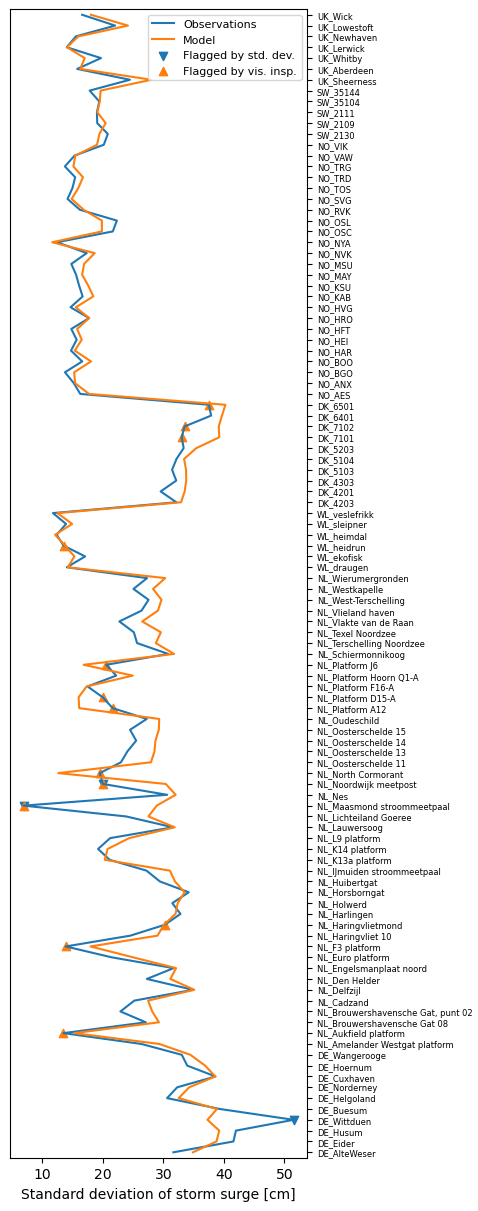}\label{fig:stddev_obs_mod}}  
  \subfloat[Hindcast - observation difference]{\includegraphics[scale=0.5]{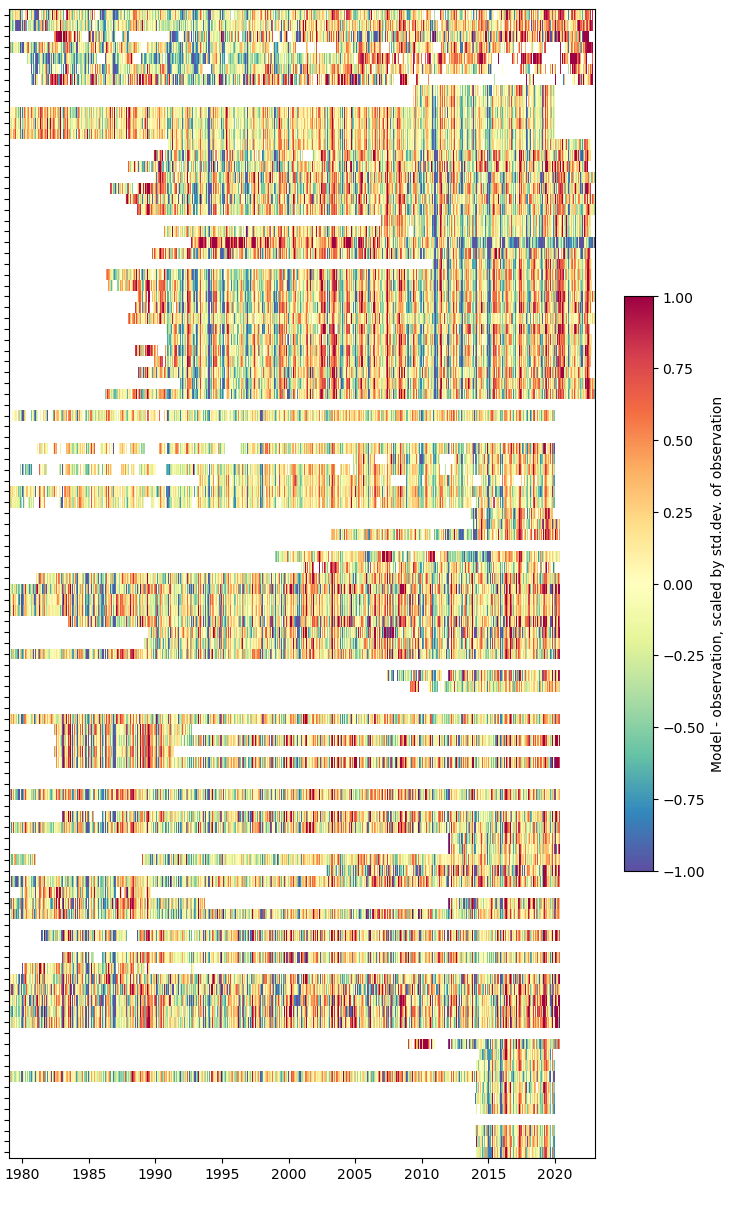}\label{fig:time_cover_diff_obs_mod}}
  \caption{\textbf{Left panel:} Comparison of standard deviation for storm surge between hindcast and observed time series for each station. In addition, the stations that were disregarded in the analysis are marked by triangles. \textbf{Right panel:} Difference between model and quality controlled observations scaled by the standard deviation for storm surge for each station for all stations for the entire hindcast period. The colorbar indicate differences in the range $\pm1$ standard deviation. Note that the temporal coverage of the observing stations varies significantly.
  Stations are grouped by country.}
  \label{fig:diff_stddev}
\end{figure}

Figure \ref{fig:stddev_obs_mod} shows a comparison of the standard deviation for the time series for both the hindcast data and the observational data for all stations. This can be viewed as a measure of storm surge activity, and clearly shows that the hindcast data exhibit similar properties with regards to activity as the observations,  and hence confirms the general quality of the hindcast. In Figure \ref{fig:time_cover_diff_obs_mod} we present the difference between the hindcast and observational data scaled by the standard deviation for storm surge for each station. This can be seen as a measure of the relative error, comparable between the different stations.

\begin{equation}
    \mbox{MAE} = \frac{\sum_{i=1}^{n}\mid \mbox{obs}_i - \mbox{mod}_i \mid}{n}
    \label{eq:mae}
\end{equation}

\begin{equation}
    \mbox{RMSE} = \sqrt{\frac{\sum_{i=1}^{n} (\mbox{obs}_i - \mbox{mod}_i)^2}{n}}
    \label{eq:rmse}
\end{equation}

\begin{equation}
    \mbox{MAE}/\sigma = \frac{\sum_{i=1}^{n}\mid \frac{\mbox{obs}_i - \mbox{mod}_i}{\sigma} \mid}{n}
    \label{eq:scaled_mae}
\end{equation}

\begin{equation}
    \mbox{RMSE}/\sigma = \sqrt{\frac{\sum_{i=1}^{n} (\frac{\mbox{obs}_i - \mbox{mod}_i}{\sigma})^2}{n}}
    \label{eq:scaled_rmse}
\end{equation}

\begin{table}[ht]
\begin{tabular}{|l|l|l|l|l|}
\hline
                          & \textbf{MAE} [cm]   & \textbf{RMSE} [cm] & \textbf{MAE/$\sigma$} & \textbf{RMSE/$\sigma$}  \\
\hline
All data                  & 9.7  & 12.4 & 0.42 & 0.54 \\
Masked less than +/- 10cm & 10.1 & 12.9 & 0.44 & 0.56 \\
Masked less than +/- 30cm & 11.8 & 15.2 & 0.52 & 0.67 \\
\hline
\end{tabular}
\caption{General values for Mean Absolute Error (MAE), Root Mean Square Error (RMSE), MAE and RMSE scaled by the standard deviation ($\sigma$) of the observed storm surge for each station (MAE/$\sigma$ and RMSE/$\sigma$) averaged over all stations over the entire hindcast period. The rows with "masked" values indicate MAE and RMSE values when masking the data for the time periods when the observed absolute values of storm surge is less than $10$ and $30$~cm. The masking thresholds removes $43\%$ and $85\%$ of the data, respectively.}
\label{tab:stats1}
\end{table}

\begin{figure}[ht]
  \centering
  \subfloat[Linear scale, all stations]{\includegraphics[width=0.5\textwidth]{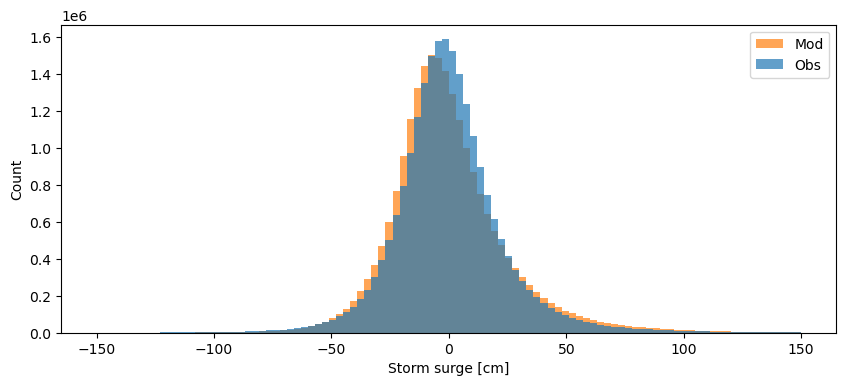}\label{fig:hist_lin}}
  \subfloat[Linear scale, Norwegian stations]{\includegraphics[width=0.5\textwidth]{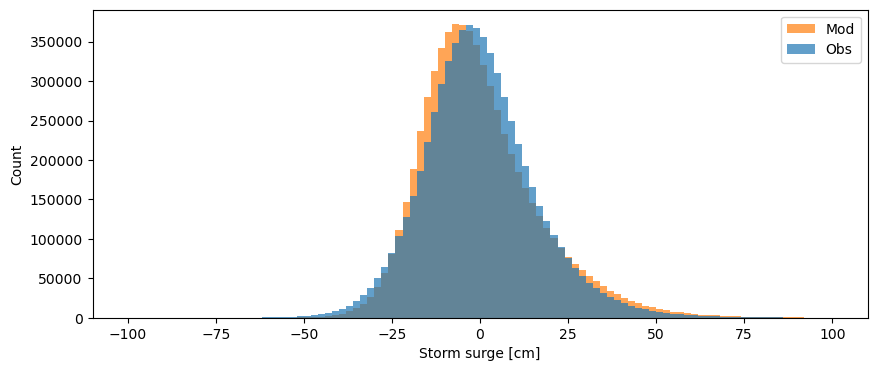}\label{fig:hist_lin_no}}
  \hfill
  \subfloat[Log scale, all stations]{\includegraphics[width=0.5\textwidth]{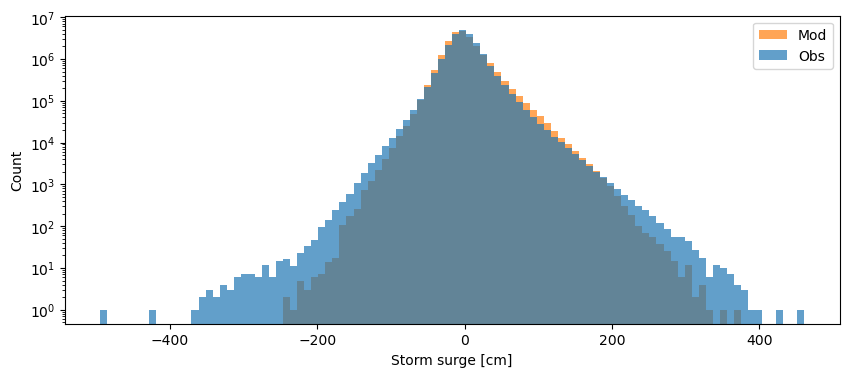}\label{fig:hist_log}}  
  \subfloat[Log scale, Norwegian stations]{\includegraphics[width=0.5\textwidth]{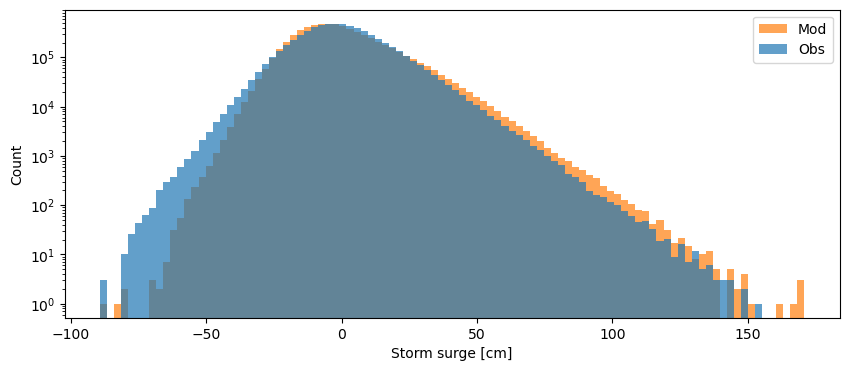}\label{fig:hist_log_no}}  
  \caption{Histograms showing the distribution of storm surge from the hindcast (Mod) and the observations (Obs) using a linear scale for the count for all stations in panel \ref{fig:hist_lin} and the Norwegian stations in panel \ref{fig:hist_lin_no}, and a logarithmic scale to further emphasize the tail of the distribution for all stations in panel \ref{fig:hist_log} and the Norwegian stations in panel \ref{fig:hist_log_no}. Note the different range of storm surge in the different figures.}
  \label{fig:hist_mod_obs}
\end{figure}

\begin{figure}[ht]
  \centering
  \subfloat[All stations]{\includegraphics[width=0.5\textwidth]{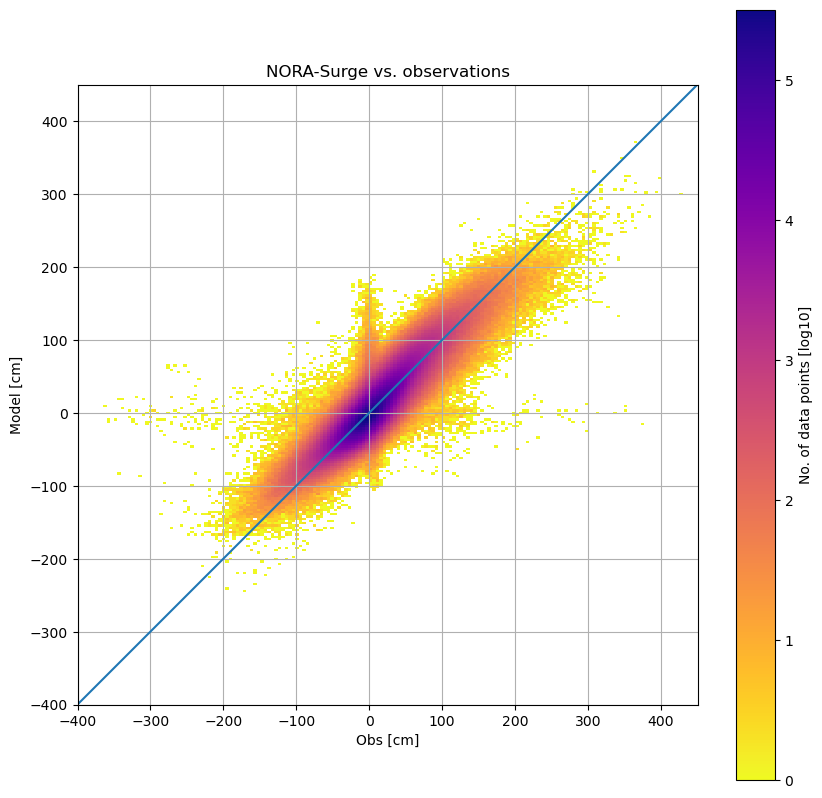}\label{fig:heatmap_obs_mod_all}}
  \subfloat[Norwegian stations]{\includegraphics[width=0.5
\textwidth]{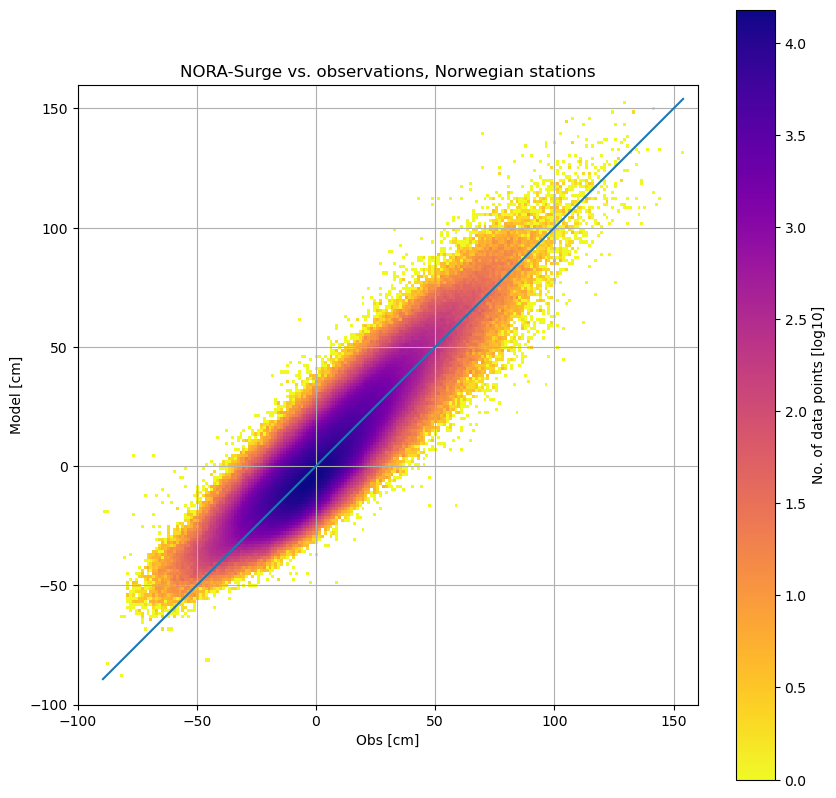}\label{fig:heatmap_obs_mod_no}}  
  \caption{Scatter plot (heatmap) showing the density distribution (on a logarithmic scale) of modeled vs observed water level for all stations (left panel) and Norwegian stations (right panel) for the entire hindcast period. Note that the axes and color scale are different for the two panels.}
  \label{fig:heatmap_obs_mod}
\end{figure}

Average values for Mean Absolute Error (MAE, see Equation \ref{eq:mae}) and Root Mean Square Error (RMSE, see Equation \ref{eq:rmse}) for all stations combined given both in cm and as fractions scaled by the standard deviation of the observed storm surge for each station (MAE/$\sigma$, see Equation \ref{eq:scaled_mae}, and RMSE/$\sigma$, see Equation \ref{eq:scaled_rmse}), averaged over the entire hindcast period, are presented in Table \ref{tab:stats1}. When including all the available quality controlled observations in the comparison with the model data we get a MAE of $9.7$~cm and RMSE of $12.4$~cm, and a scaled error for MAE/$\sigma$ of $0.42$ and RMSE/$\sigma$ of $0.54$. The value for RMSE in the current work is in line with the work by \cite{FERNANDEZMONTBLANC2020105367} where they report an RMSE for the North Sea and Norwegian sea regions of $14$~cm and $11$~cm respectively.
On average, the storm surge contribution to total water level is relatively small, as can be seen in the histograms in Figure \ref{fig:hist_mod_obs}, showing the distribution of modeled and observed water level, and in Figure \ref{fig:heatmap_obs_mod}, which displays the density scatterplot of observations vs modeled values. To investigate the events when there is actually a significant contribution from storm surge to the total water level, we have also added the values of MAE and RMSE where we have masked the data when the observed absolute value of the amplitude of storm surge is less than $10$~cm and $30$~cm in Table \ref{tab:stats1}. These thresholds remove $43\%$ and $85\%$ of the data, respectively. As expected, and as explained by \citet{kristensen22}, when the amplitudes of storm surge increase, the MAE and RMSE are expected to increase.

The distributions of modeled and observed water level in Figure \ref{fig:hist_mod_obs} indicate that the hindcast has too many occurrences of the values between $-50$ and $0$~cm and from $20$~cm upwards, and too few in the range between $0$ and $20$~cm when considering all stations combined (Figure \ref{fig:hist_lin}). For the Norwegian stations, the increased count for hindcast values is visible for the range $-25$ to $0$~cm and $25$ to $75$~cm, and the most pronounced increase in count of observed values is found between $0$ and $25$~cm (Figure \ref{fig:hist_lin_no}). For the range extending further above $+25$~cm, the hindcast has a slight increase in count for values up to about $+200$~cm for all stations. When examining the details of the tail of the distribution, for both positive and negative water level in Figures \ref{fig:hist_log} and \ref{fig:hist_log_no}, we note that the hindcast generally does not contain the same extremal values as the observations. Also, we would like to point out the four outliers with values below $-400$~cm and above $+400$~cm in Figure \ref{fig:hist_log}. These are believed to be unphysical outliers in the observational dataset that evaded our quality control.

Together with the previously mentioned non-linear interactions between e.g. tides and storm surge that are part of the observational data, but not accounted for in the model, there are other factors that can contribute to the total error described in this section. Other sources of errors include local contributions to the water level by wave setup \citep{stockdon06,melet18}, errors in the atmospheric forcing itself or how the momentum transfer from the atmosphere to the ocean is parameterized as stresses. The latter includes how momentum is transferred from the atmosphere to the ocean, and distributed in time and space, via the generation and decay of surface waves \citep{saetra2007,breivik2015}. Choices made during the model setup, like the parameterization of e.g. bottom drag, and selection of parameters can produce errors and inaccuracies. In addition, ice covered waters, where and when present, will experience a reduced momentum transfer from the atmosphere to the ocean \citep{wang2023}, which is also not accounted for in the current configuration of our hindcast. Lastly, we point out the model resolution as a possible source of error due to limited abilities to resolve features like e.g. coastline geometry and bottom topography.

The heatmaps in Figure \ref{fig:heatmap_obs_mod} display the two-dimensional distribution of corresponding hindcast and observation values.
We note a small tendency for observations to be clustered around zero in the distribution for all stations (Figure \ref{fig:heatmap_obs_mod_all}). The same is seen, although less pronounced, for the hindcast, with a slight increase in density along the $x$-axis. This could be caused by either unrealistic or erroneous observations still left in the dataset after quality control, or wrong characteristics for the hindcast for some stations/periods. The same is not seen in the distribution for the Norwegian stations (Figure \ref{fig:heatmap_obs_mod_no}). Further investigation into the subject narrows the feature down to stations located in the Netherlands (most pronounced), Germany and Denmark. Since the number of data points with this behaviour is relatively small\footnote{Further analysis show that the cases when the absolute value of the observation is below $10$~cm, and the absolute value of the model is larger than $75$~cm, account for 0.04\% of the total quality controlled observations.} (note the log scale for density), further efforts to identify and possibly remove these data points/stations has not been made. This might also be an indication that the NORA-Surge hindcast in some cases has problems capturing the exact behaviour of the storm surge for shallow stations located in the German Bight and surrounding area where tide-surge interaction might be of importance.

\begin{table}[ht]
\begin{tabular}{|l|l|l|l|l|l|}
\hline
\textbf{}                               &                                    & \textbf{MAE} [cm] & \textbf{RMSE} [cm] & \textbf{MAE/$\sigma$} & \textbf{RMSE/$\sigma$} \\ \hline
\multirow{3}{*}{\textbf{Denmark}}       & \textit{All data}                  & 10.7              & 13.7               &  0.33               &  0.42                \\ 
                                        & \textit{Masked less than +/- 10cm} & 10.9              & 14.0               &  0.33               &  0.43                \\ 
                                        & \textit{Masked less than +/- 30cm} & 12.2              & 16.0               &  0.38               &  0.49                \\ \hline
\multirow{3}{*}{\textbf{Germany}}       & \textit{All data}                  & 13.9              & 17.8               &  0.39               &  0.49                \\ 
                                        & \textit{Masked less than +/- 10cm} & 14.3              & 18.5               &  0.40               &  0.51                \\ 
                                        & \textit{Masked less than +/- 30cm} & 16.2              & 21.1               &  0.45               &  0.59                \\ \hline
\multirow{3}{*}{\textbf{Netherlands}}   & \textit{All data}                  & 11.7              & 15.0               &  0.44               &  0.57                \\ 
                                        & \textit{Masked less than +/- 10cm} & 12.0              & 15.4               &  0.46               &  0.59                \\ 
                                        & \textit{Masked less than +/- 30cm} & 13.8              & 17.8               &  0.53               &  0.68                \\ \hline
\multirow{3}{*}{\textbf{Norway}}        & \textit{All data}                  & 6.9               & 8.6                &  0.44               &  0.54                \\ 
                                        & \textit{Masked less than +/- 10cm} & 7.2               & 8.9                &  0.46               &  0.56                \\ 
                                        & \textit{Masked less than +/- 30cm} & 8.5               & 10.4               &  0.53               &  0.65                \\ \hline
\multirow{3}{*}{\textbf{Sweden}}        & \textit{All data}                  & 6.5               & 8.3                &  0.34               &  0.43                \\ 
                                        & \textit{Masked less than +/- 10cm} & 6.8               & 8.7                &  0.35               &  0.45                \\ 
                                        & \textit{Masked less than +/- 30cm} & 8.2               & 10.3               &  0.43               &  0.53                \\ \hline
\multirow{3}{*}{\textbf{UK}}            & \textit{All data}                  & 8.5               & 10.7               &  0.46               &  0.59                \\ 
                                        & \textit{Masked less than +/- 10cm} & 9.1               & 11.5               &  0.50               &  0.63                \\ 
                                        & \textit{Masked less than +/- 30cm} & 11.2              & 14.0               &  0.64               &  0.80                \\ \hline
\multirow{3}{*}{\textbf{Oil platforms}} & \textit{All data}                  & 5.7               & 7.6                &  0.42               &  0.55                \\ 
                                        & \textit{Masked less than +/- 10cm} & 6.4               & 8.7                &  0.46               &  0.63                \\ 
                                        & \textit{Masked less than +/- 30cm} & 9.9               & 14.9               &  0.72               &  1.07               \\ \hline
\end{tabular}
\caption{Same as Table \ref{tab:stats1}, but grouped by country. The group denoted by "Oil platforms" contains Norwegian offshore platforms in the North Sea (with prefix WL in e.g. Table \ref{tab:obs_source}).}
\label{tab:stats_country}
\end{table}

As previously mentioned in Section \ref{sec:model}, our hindcast model setup is based on the operational forecast model for storm surge used by the Norwegian Meteorological Institute (as described in Section \ref{sec:model}). The model has been used for many years to provide forecasts of storm surge along the Norwegian Coast. However, even if the model domain covers a significant area outside of the Norwegian Coast, little attention has previously been given to the quality of the forecast for those areas. Geographical differences in the quality of the hindcast could mainly, but not limited to, be attributed to two factors: One is the fact that storm surges behave differently in different geographical areas due to, e.g., different characteristics of bottom topography, coastline geometry and weather patterns. The other could be that the model has worse performance and quality in some regions than others. The present work is not aimed at, or sufficient for, distinguishing between the two. Some geographical differences in the behaviour/characteristics of the storm surge can be seen in Figure \ref{fig:stddev_obs_mod} where we display the standard deviation for the model and observations for all stations (grouped by country). Figure \ref{fig:time_cover_diff_obs_mod} depicts the simple difference between model and observation over time as a fraction of the standard deviation for each station (as shown in Figure \ref{fig:stddev_obs_mod}). The figure show some differences between the individual stations, and perhaps some tendency towards systematic differences between the different countries, but these are not very clear.
In Table \ref{tab:stats_country} we provide the same type of comparison as in Table \ref{tab:stats1}, but grouped by country. This emphasize that there are large variations in the quality of the hindcast when considering absolute values of MAE and RMSE, but when considering the "relative error" (MAE and RMSE based on differences scaled by the standard deviation) the geographical differences get less pronounced. The scaled values for MAE range from $0.33$ to $0.46$ and RMSE from $0.42$ to $0.59$ standard deviations.
The smallest absolute errors are found at the stations in Sweden, Norway and the offshore oil platforms located in the North Sea, with values of MAE and RMSE, respectively, of less than $7$~cm and $9$~cm. This is less than the averages over all stations, and  more in line with the values of MAE and RMSE of $3$ and $5$~cm for the errors reported at initial time ($+0$~h forecast lead time) by \cite{kristensen22} for the Norwegian storm surge forecast system evaluated at Norwegian stations. However, we would like to point out that the forecasts used in the evaluation by \citet{kristensen22} are post-processed and subject to correction by using the difference between model and observations, and hence exhibit lower error values than the free run in the NORA-Surge hindcast. The purpose of such corrections is to reduce low frequency error components in the reference level for the water level present in the operational forecast for a discrete set of stations on the Norwegian Coast. The correction does not affect the amplitude of storm surges, and does not correct the two-dimensional model field. Thus, such a correction is not applicable along the whole coastline in the present hindcast.

\begin{figure}[ht]
    \centering
    \includegraphics[width=0.85\textwidth]{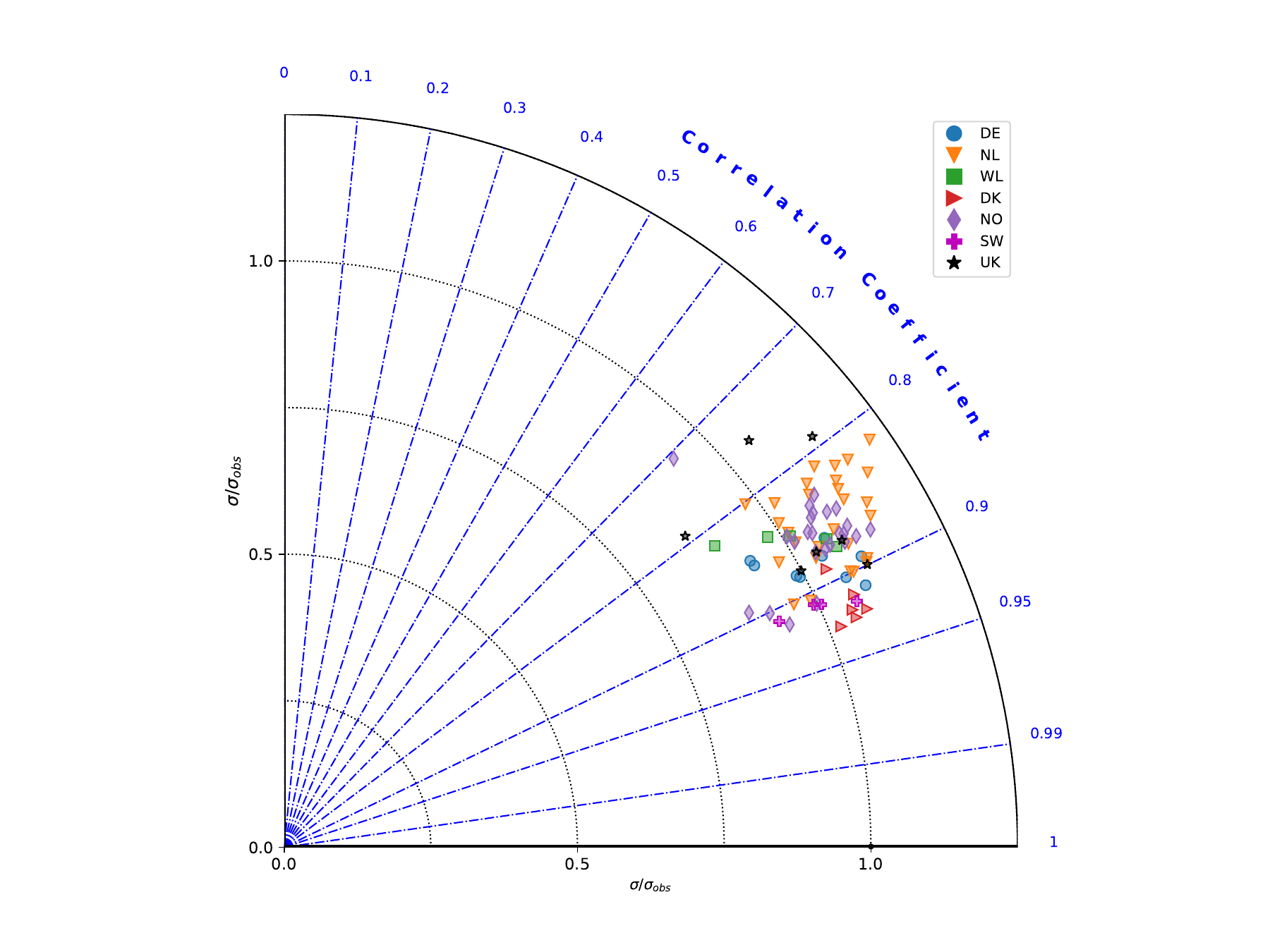}
    \caption{Taylor diagram showing the normalized error statistics for all the quality controlled stations in the NORA-Surge hindcast compared to observations. Each marker in the plot represents a station, and different colors and symbols are used to identify which country they belong to. The model's variance has been normalized with respect to the observed variance.}
    \label{fig:taylor_diagram}
\end{figure}

The Taylor diagram in Figure \ref{fig:taylor_diagram} shows normalized error statistics for each station in the NORA-Surge hindcast compared to measurements. We see that the relative errors in the hindcast do not vary much between the different stations, with correlation coefficients mainly in the interval $0.80-0.95$, and normalized standard deviations in the interval $0.85-1.25$. However, there are some minor geographical differences, and stations from the same region tend to cluster into smaller groups.

We note two outliers in the Taylor diagram, the top left NO station with a correlation of approximately $0.7$ and the UK station with a correlation of $0.75$. The NO station is Ny-Ålesund (NO\_NYA) located on the Arctic island of Spitsbergen, and the UK station is Newhaven (UK\_Newhaven), located just west of Dover (in the eastern part of the English Channel). 
For the NO\_NYA station, we note that there seems to be a negative trend in the difference between hindcast and observation in Figure \ref{fig:time_cover_diff_obs_mod}. This trend is explained by the rapid vertical land uplift at Svalbard \citep{kierulf2021}. 
The UK\_Newhaven station is placed near the south-western boundary of the hindcast model domain, and the fact that it is located west of Dover means it will be subject to storm surges that enter the English Channel from the west. The lack of proper boundary values from an outer model, as mentioned in Section \ref{sec:model}, could cause some of the storm surge signal that should have entered from the North Atlantic to be missing in the hindcast for this area of the English Channel. 

The two UK stations with correlation just below $0.8$ could also be considered as outliers since they deviate from the rest of the UK cluster. These are the stations UK\_Whitby and UK\_Sheerness, which are both located at or near river mouths. River runoff can cause local elevation changes that are not captured by the hindcast.

\begin{figure}[ht]
  \centering
  \subfloat[All stations]{\includegraphics[width=0.5\textwidth]{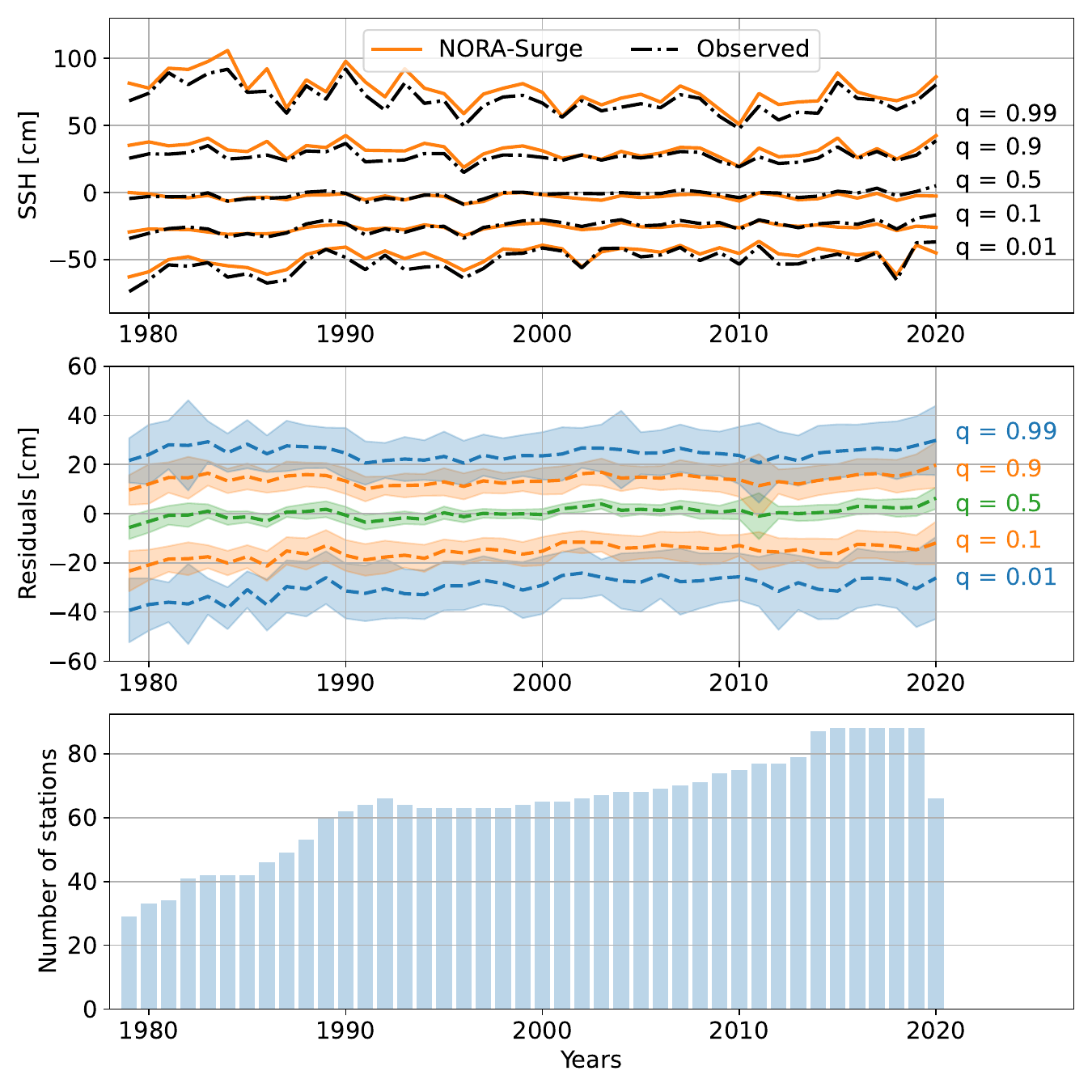}\label{quantiles_all_stations}}
  \subfloat[Norwegian stations]{\includegraphics[width=0.5
\textwidth]{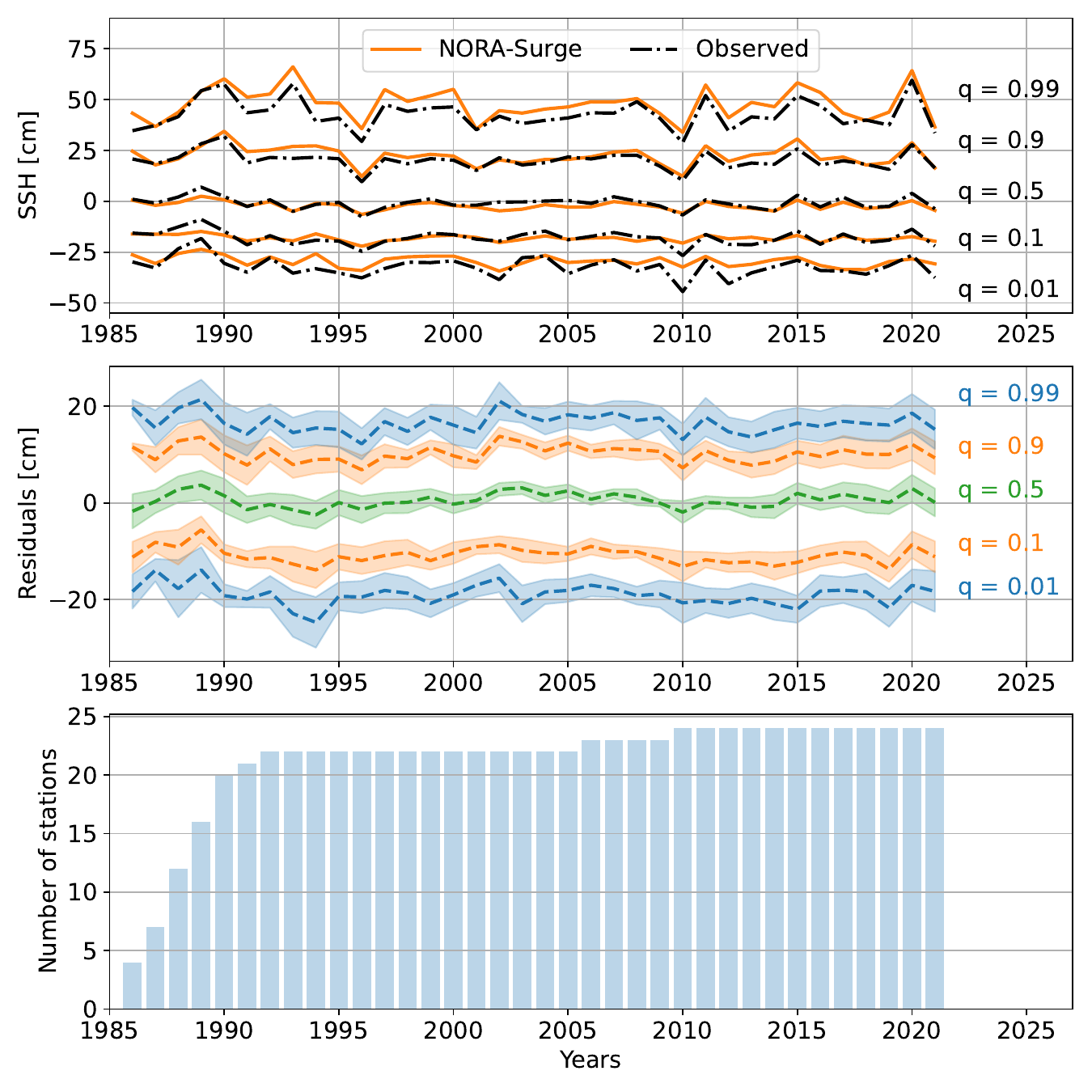}\label{quantiles_no}}  
  \caption{The upper panels show the annual quantiles of observed SSH (dashed-dotted black line) and NORA-Surge (solid orange line) averaged over the stations. In the middle, the average annual quantiles of the residuals (observed - NORA-Surge) are plotted (dashed lines). The shadows represent the uncertainties in terms of 1 standard deviation. The lower panels show the number of stations with observations per year. The figures in (a) have been generated with data from all the stations that passed the quality control, while the figures in (b) were generated with data from the Norwegian stations.}
  \label{fig:quantiles_ssh_uncertainties_no_stations_v2}
\end{figure}

Figure \ref{fig:quantiles_ssh_uncertainties_no_stations_v2} shows the quantiles of annual SSH and residuals averaged over all stations in Figure \ref{quantiles_all_stations}, and only the Norwegian stations in Figure \ref{quantiles_no}. This confirms our previous observations: The hindcast is in good agreement with the observations, not only for average values but also in the tails of the distributions. Although, for the highest and lowest quantiles showed in this figure, the modelled SSH (solid orange line) is generally mostly higher than the observed values (dashed-dotted black line), this result cannot be extrapolated to the full range of SSH (as shown in Figure \ref{fig:hist_mod_obs} and \ref{fig:heatmap_obs_mod}).

The SSH median values has very little variability around zero (see the upper panels), for both observed and model values, but extreme values have more interannual variability, as there are more events some years compared to others. The extremes are also sensitive to the number of stations (shown in the bar plots in the bottom panels and in Figure \ref{fig:time_cover_diff_obs_mod}), in particular before 1990, when we have less observations. In addition, the distributions are not symmetric (see also Figure \ref{fig:hist_mod_obs} and \ref{fig:heatmap_obs_mod}), and positive SSH values are in general larger, in absolute value, than negative values. 

The panels in the middle show the average annual quantiles of the residuals (the difference between observed and modelled surges) $\pm 1$ standard deviation, depicting the variability between stations. We see that the quantiles of the residuals depend on both time and location. And the further we move to the tails, the larger the variability across stations. There is a slight indication of a trend in the median computed for all stations. This could be explained by a consistent trend in the measurements due to e.g. sea level rise, but these are speculations that have not been investigated. 

If we compare Figure \ref{quantiles_all_stations} with Figure \ref{quantiles_no} (note that the scale of the vertical axes is different), we see that the quantiles, of both SSH and residuals, are greater in absolute value when computed for all the stations than only for the Norwegian stations. The same yields for the standard deviation on the annual residuals. This is due to the fact that storm surge generally has lower amplitudes for Norwegian stations than the average over all stations, and is in agreement with e.g. the standard deviation for storm surge as shown in Figure \ref{fig:stddev_obs_mod} and the distribution of storm surge as shown in Figure \ref{fig:hist_mod_obs}.

Finally, when analyzing Figure \ref{fig:quantiles_ssh_uncertainties_no_stations_v2}, it is important to remember that the results have been aggregated for an heterogeneous dataset with stations located all around the North Sea, and that these locations experience typical surges of different magnitudes. The average quantiles have been computed for absolute values, meaning that they are not representative of one particular station. Therefore, the relative errors for each station are better represented in the Taylor diagram in Figure \ref{fig:taylor_diagram}.

\subsection{Model statistics and extreme value estimates of water level}\label{sec:stat_extremes}
As mentioned in Section \ref{sec:introduction}, one of the main reasons for producing hindcasts of storm surges, such as the NORA-Surge hindcast, is to gain knowledge of the statistical distribution of the storm surge, and assess water level extremes in coastal regions. Here we present an overview of the statistics of the storm surge hindcast together with an extreme value analysis of 100-year return values for maxima and minima. Good knowledge about return periods and extreme values for storm surge is of great importance for planning and utilization of the coastal zones that are subject to influence of high water levels.
In a changing climate, this becomes even more important than before, as sea level rises and new areas become susceptible to such influence. One could argue that such statistics could be obtained by analyzing observational data from water level stations alone, but this would not provide sufficient geographical coverage for all areas since water level stations generally does not have a very high coverage along the coast everywhere, as shown in Figure \ref{fig:obs_map}. And as shown in Figure \ref{fig:time_cover_diff_obs_mod}, the time coverage of the available stations with observations varies. The use of models to produce hindcasts is therefore a cost effective way to fill the gaps between observational sites, and can also provide longer time coverage than observations. Hindcast datasets are limited in time, space and quality by the available forcing data, such as the NORA3 hindcast, and the available compute resources.

\begin{figure}[!ht]
  \centering
  \subfloat[]{\includegraphics[width=0.6\textwidth]{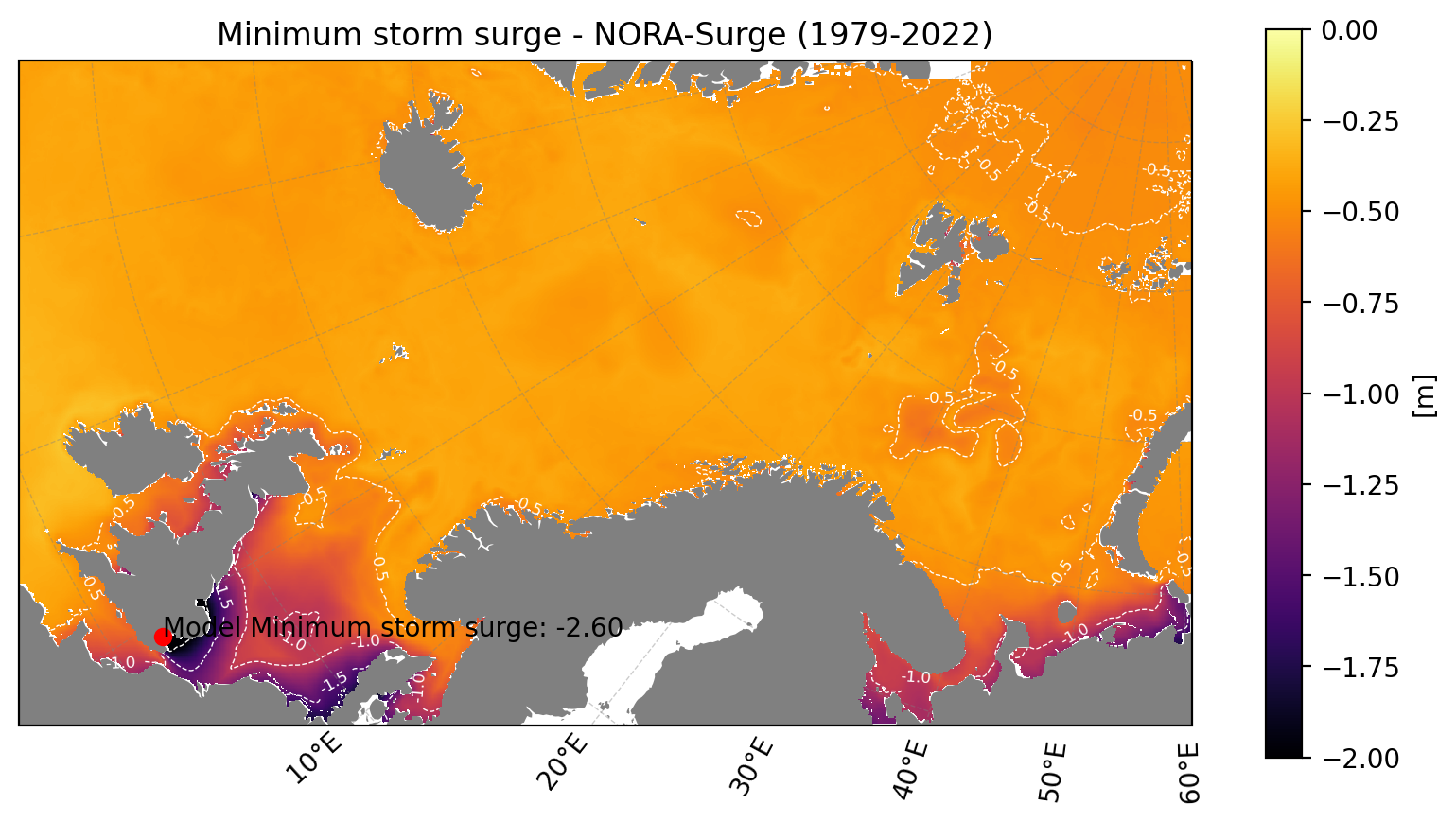}\label{fig:zeta_min}}
  \hfill
  \subfloat[]{\includegraphics[width=0.6\textwidth]{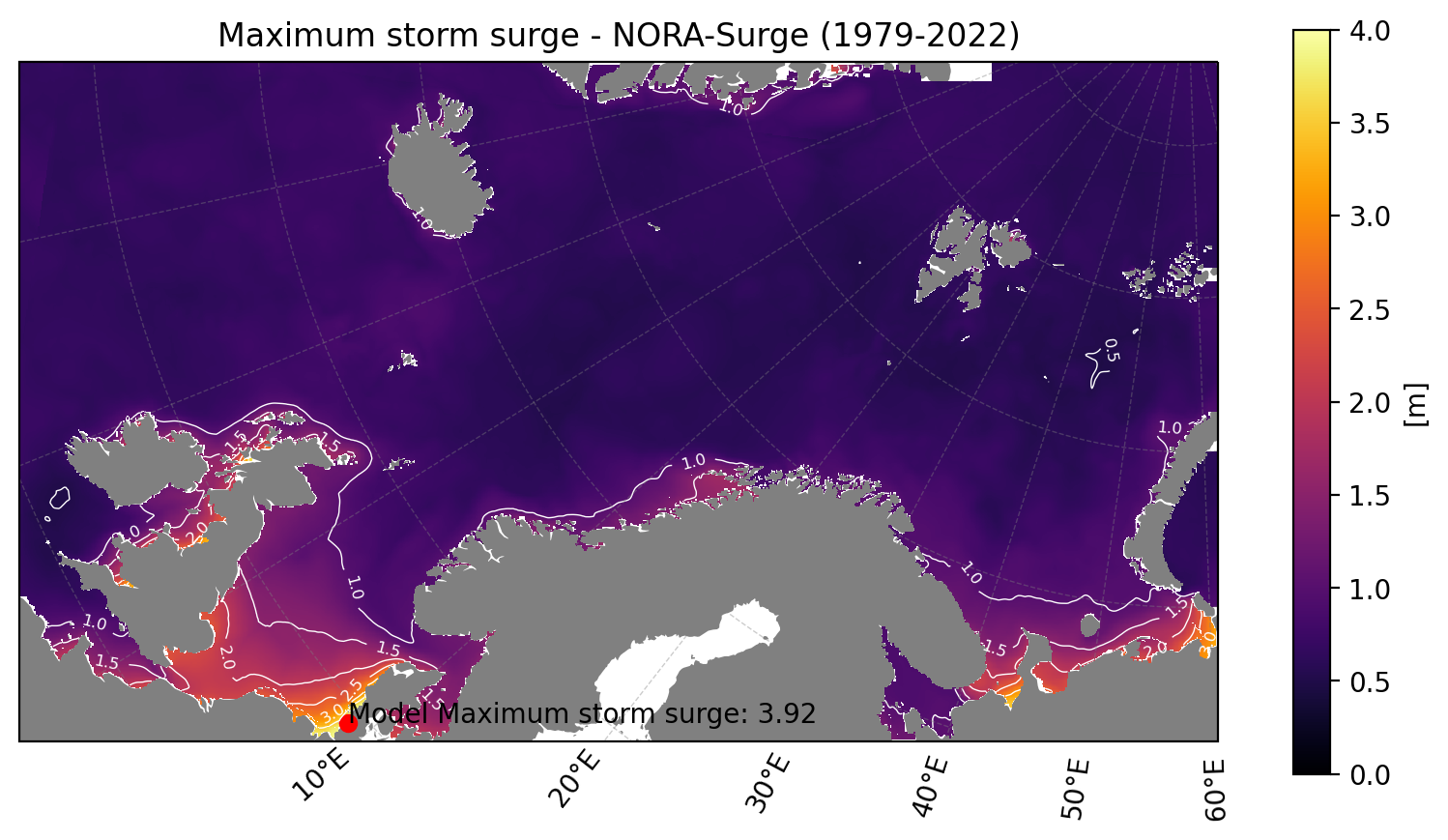}\label{fig:zeta_max}}  
  \hfill
  \subfloat[]{\includegraphics[width=0.6\textwidth]{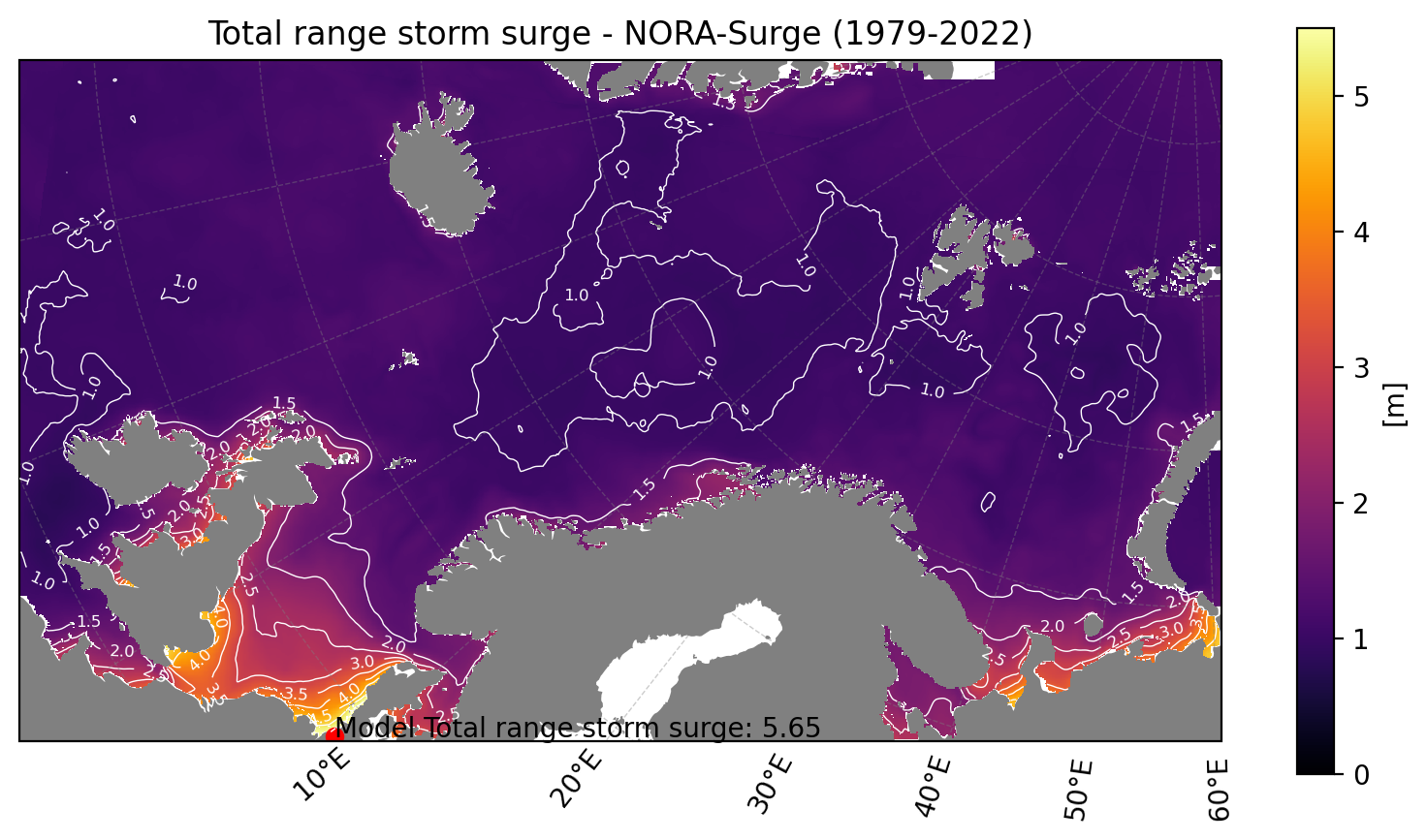}\label{fig:zeta_range}}  
  \caption{The minimum (\ref{fig:zeta_min}), maximum (\ref{fig:zeta_max}) and range (\ref{fig:zeta_range}) of storm surge water level found throughout the NORA-Surge hindcast (1979--2022).}
  \label{fig:nora-surge_stats}
\end{figure}

The areas that are most influenced by storm surges in the area covered by our hindcast are the coast of the United Kingdom, the Netherlands, the West Coast of Denmark, the area around Lofoten in Norway and the Russian coast between the Kola Peninsula and Novaya Zemlya. Generally, as shown in Figure \ref{fig:nora-surge_stats}, these areas contain both the highest and lowest values of storm surge in the hindcast period, and hence also the largest ranges of storm surge.
The largest minimum value for storm surge in the hindcast, as can be seen in Figure \ref{fig:zeta_min}, is $-2.60~m$, and is found on the south eastern coast of the UK, near the mouth of the river Thames.
The largest value for maximum storm surge, as shown in Figure \ref{fig:zeta_max}, is $3.92~m$, and is located in the German Bight, near the mouth of the river Elbe close to Hamburg (Germany). This area is a hot spot for storm surges, and also includes the largest storm surge range in the hindcast of $5.65~m$, as seen in Figure \ref{fig:zeta_range}. Generally, the southeastern part of the North Sea is an area of large storm surge activity, as is also illustrated by the large values of standard deviation for the DE, NL and DK stations in Figure \ref{fig:stddev_obs_mod}.

\begin{figure}[ht]
  \centering
  \subfloat[]{\includegraphics[width=0.6\textwidth]{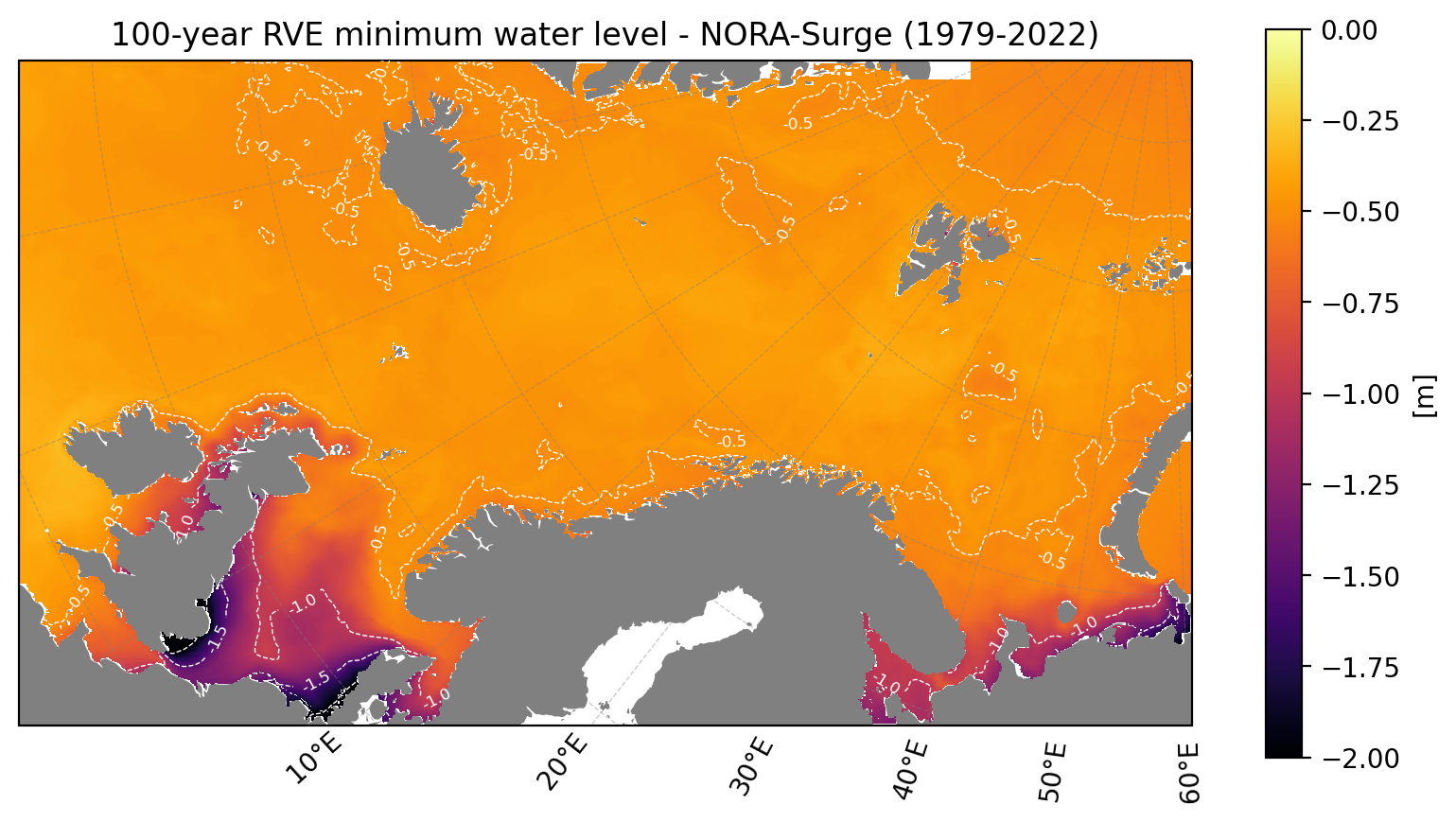}\label{fig:zeta_min100}}
  \hfill
  \subfloat[]{\includegraphics[width=0.6\textwidth]{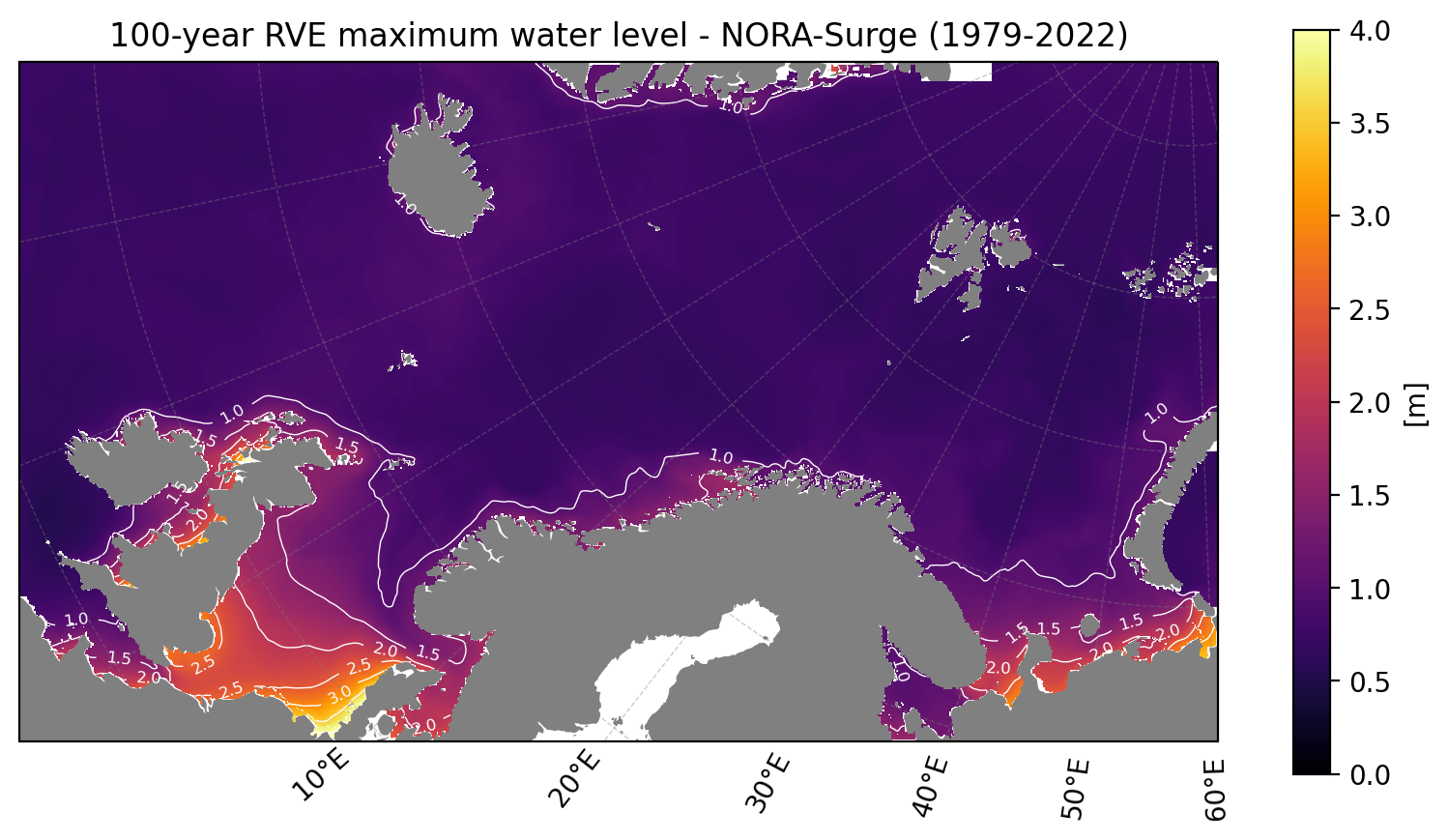}\label{fig:zeta_max100}}  
  \caption{The 100-year return value estimate of the minimum (\ref{fig:zeta_min100}) and maximum (\ref{fig:zeta_max100}) storm surge water level based on the NORA-Surge hindcast (1979--2022).}
  \label{fig:nora-surge_stats100}
\end{figure}

A 43-year long hindcast is not sufficiently long to provide robust statistics of extreme water levels by direct analysis and extraction of extremes from the model results. We have therefore performed a statistical analysis based on annual (block) maxima of the hindcast dataset and calculated the 100-year Return Value Estimates (RVE) of minimum and maximum storm surge for the model domain by fitting a Gumbel distribution \citep{gumbel1958statistics} to the annual minima and maxima, respectively. 
As shown by \citet{col01}, the cumulative distribution function (CDF) of the block maxima formed from a random sequence of independent variables and identically distributed (IID) variables will follow the Generalized Extreme Value (GEV) distribution,
\begin{equation}
   G(z) = \exp \left\{ -\left[1+\xi
   \left(\frac{z-\mu}{\sigma}\right)\right]^{-1/\xi}\right\}.
   \label{eq:gev}
\end{equation}
Here $\sigma$ is the scale parameter, $\mu$ is the location parameter, and $\xi$ is the shape parameter. The GEV distribution contains as a special case the Gumbel ($\xi = 0$)  distribution. The Gumbel distribution provides a somewhat smoother estimate than applying the the full GEV distribution, which is more sensitive to outliers. Here, we apply the maximum likelihood approach when fitting the distribution and we utilize the negative of the annual minima when estimating minimum water levels. The 100-year RVE of minima and maxima are shown in Figure \ref{fig:zeta_min100} and \ref{fig:zeta_max100}, respectively.

The overall picture for 100-year RVE is quite similar to those of minimum and maximum levels of storm surge in the hindcast, but in general yields somewhat larger absolute values. The minimum 100-year RVE is estimated to be $-2.75$~m, located in the same position as the hindcast minimum value. The maximum 100-year RVE is estimated to be $4.45$~m located on the western coast of Scotland. However, we believe this to be an artifact due to unfortunate effects of coastline geometry since the maximum values are found furthest into a fjord that generally does not contain too large values, but where the variability is very large. The maximum 100-year RVE found in the German Bight, near the position of the maximum storm surge level in the hindcast, is approximately $3.98$~m. We interpret this as a better estimate of the maximum 100-year RVE.

\subsection{Case studies}\label{sec:case_studies}
To further assess the performance of the model and its ability to reconstruct past events, we have investigated three well-known storm surge events. 
Even if these three selected cases could be viewed as extreme in a statistical sense, since all of them brought with them record high water levels, they are caused by weather conditions that are not uncommon for the part of Northern Europe covered by the NORA-Surge hindcast.

\subsubsection{The 1987 Oslofjord storm surge}\label{sec:1987_oslofjord}
On 16-17 October 1987, the southern parts of Norway were hit by a big low pressure system that passed by the western coast of Norway from south to north, see \cite{engen:1988} and \cite{andresen:etal:1987}. The weather system brought with it gale and storm force winds from the south, together with very large amounts of rain. The combination of spring tide and high storm surge resulted in the second highest water level in the Oslofjord ever recorded. This, together with river flooding due to extreme rainfall, resulted in flooding of large areas along the coast of the Oslofjord.

\begin{figure}[ht]
    \centering
    \includegraphics[width=0.75\linewidth]{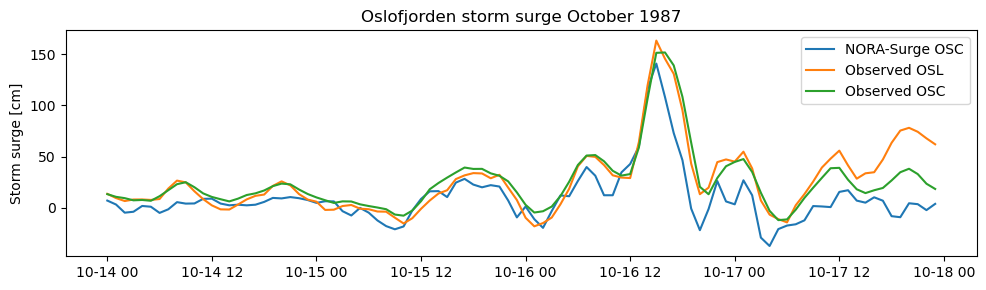}
    \caption{Time series for observed storm surge in Oslo (OSL) and at Oscarsborg (OSC) in the Oslofjord together with hindcasted storm surge at Oscarsborg from NORA-Surge. Values are given in centimeters and referenced to the Mean Sea Level.}
    \label{fig:oslofj1987}
\end{figure}

The hindcast closely follows the observed surge, although the absolute value from the hindcast is a bit low. As shown in Figure \ref{fig:oslofj1987}, the maximum observed storm surge in Oslo was $163$~cm and Oscarsborg $152$~cm. From the NORA-Surge hindcast, the maximum surge for the Oscarsborg station was $141$~cm. We do not include the time series and the maximum value for the Oslo station from the hindcast, since this is the same model grid point as Oscarsborg. This illustrates how model resolution can act as a limiting factor when modeling storm surge along a complex coastline with long and narrow fjords like the Norwegian coast and the Oslofjord. The direct comparison for the Oscarsborg station has an error for the maximum surge of $11$~cm, which amounts to $7.2\%$ of the observed storm surge peak.

\subsubsection{2013: Storm Xaver}\label{sec:2013_xaver}
\begin{figure}[ht]
    \centering
    \includegraphics[width=0.5\linewidth]{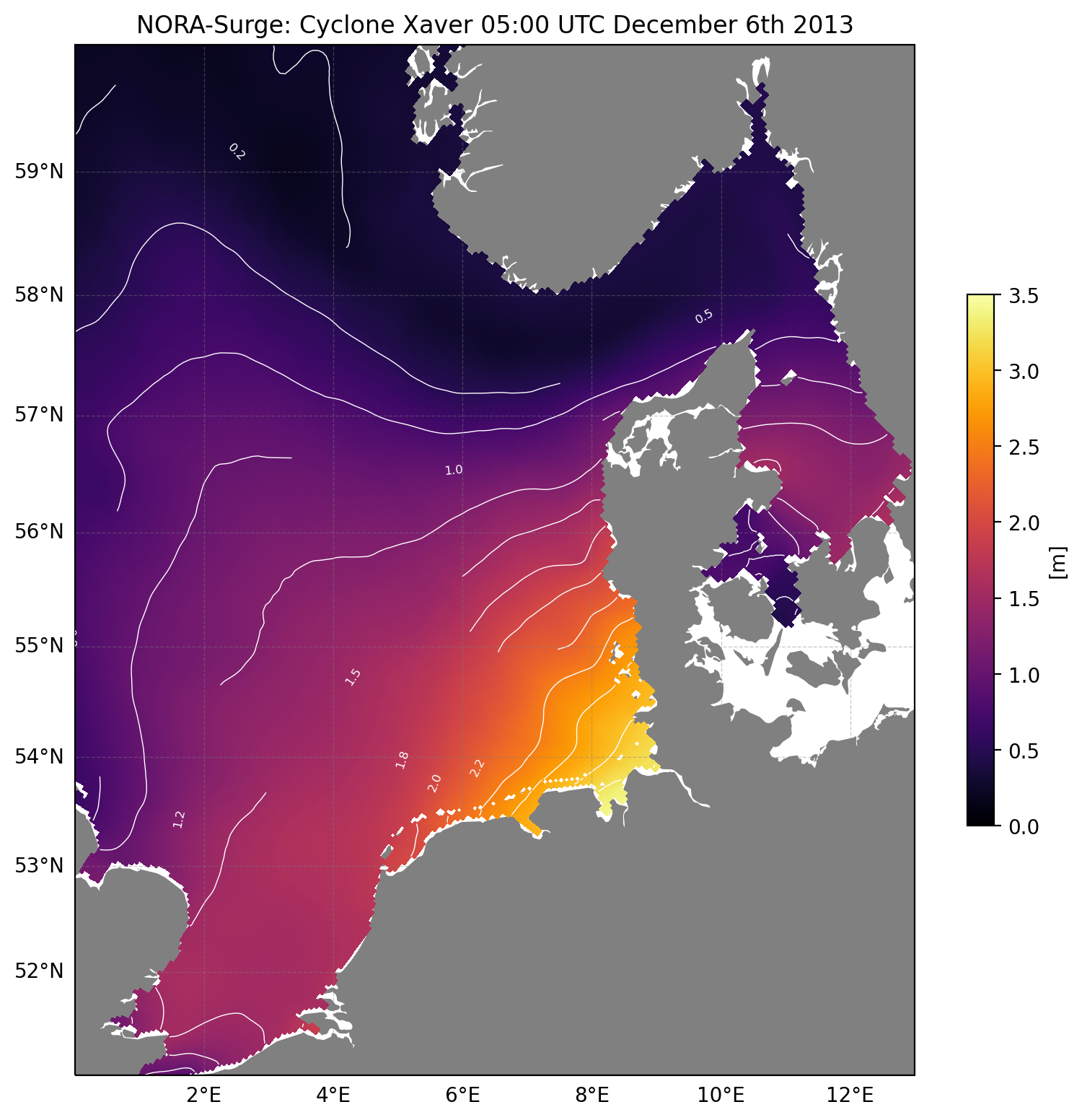}
    \caption{The NORA-Surge hindcast storm surge in the German Bight at the peak of the Xaver storm at 05 UTC on the 6th of December 2013, with a maximum value of $339$~cm.}
    \label{fig:zeta_xaver}
\end{figure}

The storm Xaver moved slowly from west to east across Northern Europe during 5-7 December 2013 \citep{deutschlander:etal:2013}. The strong northerly winds over the North Sea resulted initially in a convergence of water along the eastern coast of the United Kingdom due to Ekman transport. When the wind direction started to back left, this water started to move anti-clockwise along the North Sea coastline as a free Kelvin wave. The combined Kelvin wave with local amplifications due to Ekman transport by the wind, and direct effect of the low pressure resulted in storm surge among the top five highest recordings over the last 100 years in the German Bight according to \cite{deutschlander:etal:2013}. The storm surge in the North Sea during the peak of the storm is shown in the map in Figure \ref{fig:zeta_xaver}. 

\begin{figure}[ht]
    \centering
    \includegraphics[width=0.75\linewidth]{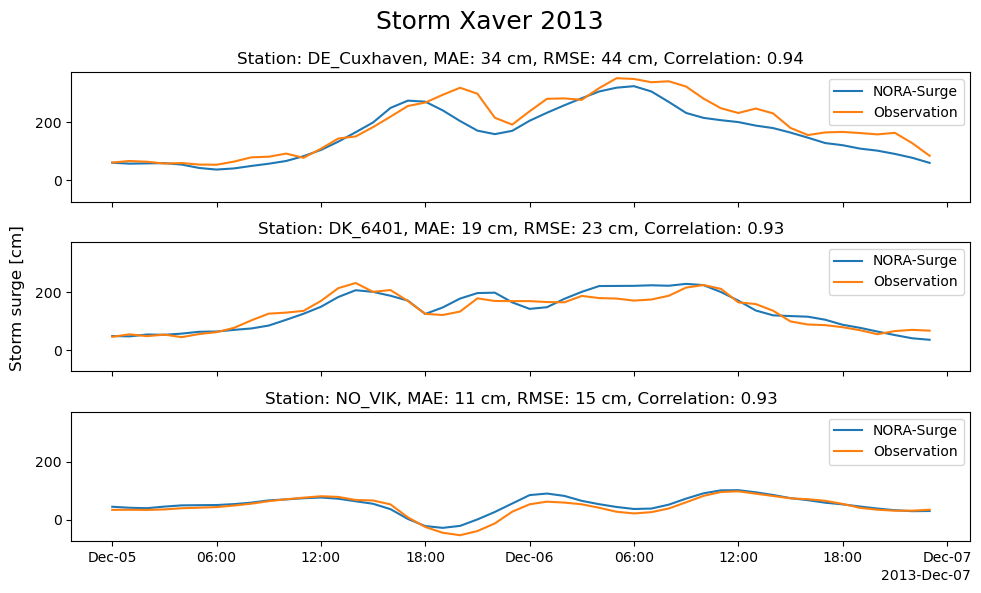}
    \caption{Time series comparison between hindcast and observation for the water level stations Cuxhaven in Germany, Esbjerg in Denmark (DK\_6401) and Viker in Norway (NO\_VIK) during the storm Xaver. Values are given in centimeters and referenced to the Mean Sea Level.}
    \label{fig:zeta_xaver_tseries}
\end{figure}

As shown in Figure \ref{fig:zeta_xaver_tseries}, the maximum observed storm surge in Cuxhaven was $352$~cm, and from the NORA-Surge hindcast, the corresponding maximum value is $325$~cm. The hindcast underestimation of $27$~cm corresponds to $7.7\%$ of the observed surge. And, as seen in Figure \ref{fig:hist_log}, these values should be considered to be extreme in a statistical sense since they are only found near the end of the tail of the statistical distribution, in the range where the model has a known underestimation. We also speculate whether some of the difference between hindcast and observations at Cuxhaven could be attributed to the non-linear interaction effects of tide-surge interaction (see \cite{horsburgh:2007} and \cite{idier:2019}). This effect can modify the phase and amplitude of the tide in shallow areas when there is a large storm surge amplitude, and hence the tidal prediction used to subtract from the observation of total water level to obtain the observation of storm surge could be inaccurate for the time period in question. Since the NORA-Surge hindcast does not contain tides, such non-linear interactions are not accounted for in the hindcast. Also, as mentioned in Section \ref{sec:obs_vs_model}, we acknowledge that due to the combination of complex coastline geometry, shallow bottom topography and large storm surge amplitudes in the southern North Sea and the German Bight, the hindcast performance in this area is less than that for the areas further north (e.g. for the Norwegian coast). This could also be a factor in explaining the difference between hindcast and observation in the Xaver case.

As the low pressure system moved further east, and the wind direction continued backing further left, the storm surge signal moved further through the North Sea up along the western coast of Denmark and into the Skagerrak area as a Kelvin wave (see Figure \ref{fig:zeta_xaver_tseries}). Even though not directly affected by the weather system itself, the Oslofjord experienced large fluctuations in water level due to the Kelvin wave, with variations in water level due to storm surge in the inner part of the Oslofjord in the range of up to $\pm1~m$ in less than 24 hours. This can be seen in the lower panel in Figure \ref{fig:zeta_xaver_tseries}, displaying the observed and modeled storm surge for the station Viker (NO\_VIK) at the entrance to the Oslofjord. Both the amplitude of the surge, and the values of MAE and RMSE, is significantly reduced compared to the two panels above for Cuxhaven and Esbjerg (DK\_6401). Correlation between hindcast and observations for the three stations for the time period shown is almost identical at $0.93$ and $0.94$. 

\subsubsection{January and February 2020: Storm surge events on the western coast of Norway}\label{sec:2019-2020_westcoast}
In January and February 2020, the western part of Norway experienced a series of extreme storm surge events. When the total water level exceeds certain thresholds for the permanent water level stations in Norway, MET Norway issues a yellow, orange or red warning to the public (see \cite{kristensen22}). In the case of a red warning, even if the only parameter exceeding the red criterion is the total water level, the event is categorized as an extreme weather event and is given a name. 

\begin{figure}[ht]
    \centering
    \includegraphics[width=0.75\linewidth]{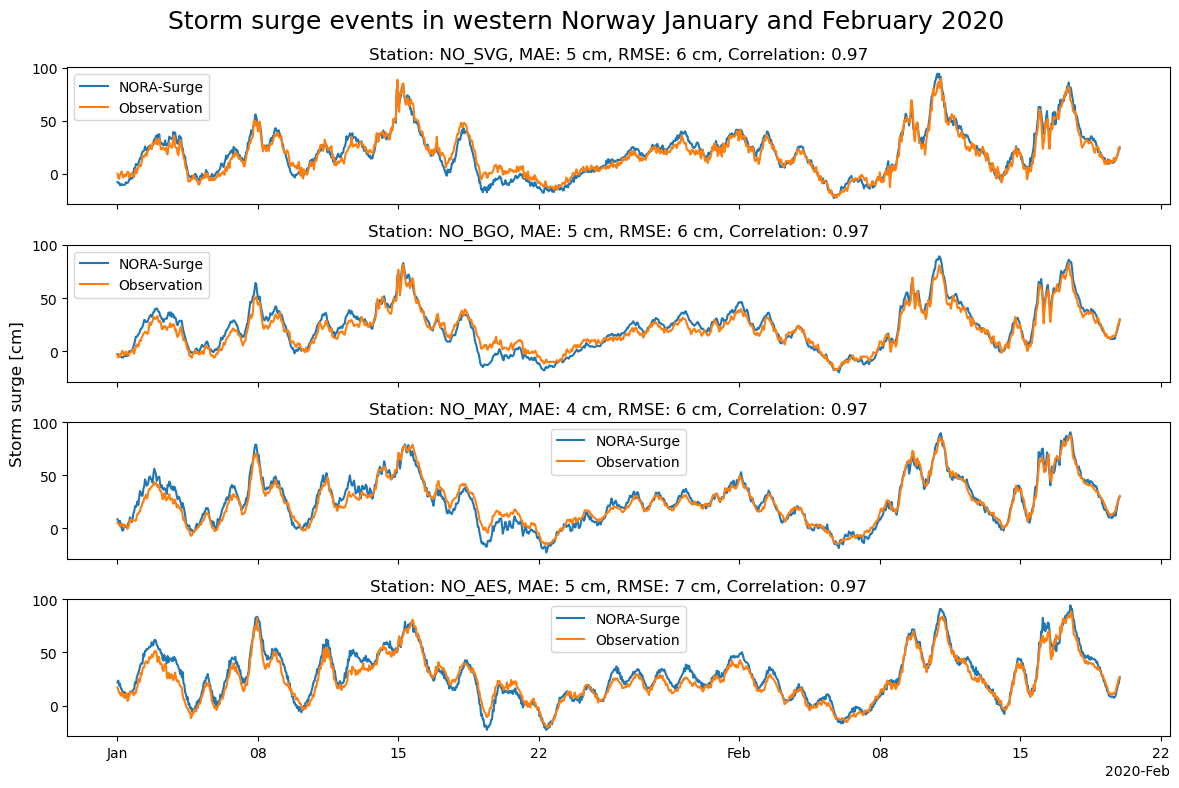}
    \caption{Time series plots of observed and hindcast storm surge between January 1st and February 20th 2020 for four selected stations on the western coast of Norway. From top to bottom the stations are Stavanger (NO\_SVG), Bergen (NO\_BGO), Måløy (NO\_MAY) and Ålesund (NO\_AES). Values for MAE, RMSE and correlation for the depicted time period for each station are shown in the title in each panel.}
    \label{fig:w_norway_jan-feb_2020}
\end{figure}

The extreme weather "Didrik" in January 2020 \citep{selberg:etal:2020} and "Elsa" in February 2020 \citep{skjerdal:etal:2020} were quite similar weather events where a deep low pressure system of around $945$~hPa passed south of Iceland and eventually moved north up along the Norwegian coast. The low pressure systems brought with them strong westerly to south-westerly winds in the northern North Sea, which in addition to the inverse barometric effect itself, resulted in a large transport of water towards the western Norwegian coast. These events resulted in a total water level among the top 10 highest observed water levels for the stations from Stavanger in the south up to Rørvik in the north. For the station in Måløy, the water level on February 11th 2020, was the highest ever recorded. Time series plots of observed and hindcast storm surge for a few selected stations on the west coast of Norway for the time period between January 1st and February 20th 2020 are shown in Figure \ref{fig:w_norway_jan-feb_2020}.

The difference between the hindcast and observed maximum values of storm surge during "Elsa" for the four selected stations are from $-6$ to $2$~cm, and during "Didrik" between $4$ and $8$~cm. Correlation for all stations are $0.97$ and RMSE values are in the range $6-7$~cm for the entire time period considered.

\section{Summary and concluding remarks}\label{sec:discussion_and_conclusion}
The NORA-Surge hindcast is an open and freely available storm surge hindcast\footnote{Available from \url{https://thredds.met.no/thredds/catalog/stormrisk/catalog.html}.} based on a barotropic setup of the ROMS ocean model with $4$~km horizontal resolution covering the areas of the North Sea, the Norwegian Sea and the Barents Sea. It includes the entire coast of the United Kingdom, the Netherlands and Norway (including Svalbard), in addition there is partial coverage of the coast of Denmark and western coast of Sweden. The temporal coverage of the hindcast is more than four decades, covering the time period from 1979 throughout 2022.

The hindcast has been extensively validated against an observational dataset of quality checked observations of water level from more than 90 coastal and offshore locations, as described in Section \ref{sec:obs_dataset}. Comparison of the hindcast with all the observational data yields average values of MAE and RMSE of $9.7$ and $12.4$~cm respectively, scaled values for MAE/$\sigma$ of $0.42$ and RMSE/$\sigma$ of $0.54$ standard deviations (see Table \ref{tab:stats1} and \ref{tab:stats_country}), and correlation coefficients for each station primarily in the range of $0.80-0.93$ (see Figure \ref{fig:taylor_diagram}). The absolute error varies between the stations, and has larger values for some geographical areas than others. The areas with the largest amplitudes in storm surge are also associated with the largest errors (see Figure \ref{fig:diff_stddev} and Table \ref{tab:stats_country}). However, as shown in the Taylor diagram in Figure \ref{fig:taylor_diagram}, the relative errors for the stations (with a few exceptions) are quite similar. There is a slight clustering of the stations from different countries, which reflects geographical and bathymetry specificities of the coastline depending on the location, but all the clusters have at least some overlap with each others.

Statistics of extreme values of minimum, maximum and range for storm surge within the hindcast are calculated and presented in the maps in  Figure \ref{fig:nora-surge_stats}. The minimum storm surge value within the hindcast is $-2.60$~m, the maximum is $3.92$~m and the largest range of storm surge is $5.65$~m. All of which are found in the southern part of the North Sea. Also, a return value estimation using the GEV distribution has been carried out based on the hindcast dataset, and the 100-year return value estimates of minima and maxima of water level for the entire hindcast model domain are presented in the maps in Figure \ref{fig:nora-surge_stats100}. The minimum and maximum return values for storm surge are $-2.75$~m and $3.98$~m, respectively, also located in the southern North Sea.

Due to the fact that the model setup used to produce the NORA-Surge hindcast does not have nesting values from an outer, preferably global, storm surge model, there is no response to the annual change in domain average surface pressure (i.e., only gradients in the surface pressure produce changes in water level). This is evident in Figure \ref{fig:time_cover_diff_obs_mod} when considering the difference between model and observations: There is a consistent annual variation in the difference. This is particularly evident when examining the Norwegian, Swedish and Danish stations.
Moreover, there is no tidal forcing in the hindcast, resulting in the absence of non-linear interaction between tide and surge, so-called tide-surge interactions. This could explain some of the errors seen in the regions of both large tidal and storm surge amplitudes in the southern part of the North Sea (e.g. the Xaver case study in Section \ref{sec:2013_xaver}).

The hindcast generally validates well against a broad range of water level measurements in the North Sea. However, despite mostly capturing the range and also upper percentiles, a number of improvements can be made. As has been mentioned, increased horizontal resolution with more detailed bottom bathymetry and a more realistic coastline, would likely reduce some errors in the hindcast. In addition, the inclusion of tides should be considered to improve the quality in regions with large tidal amplitudes and shallow water depths where surge-tide interaction is expected to be important. In very shallow areas, wave setup can also add to the water level in storm conditions \citep{stockdon06,melet18}. Finally, we expect the atmospheric forcing in the storm surge model could be improved by including wave effects in the atmospheric momentum fluxes \citep{staneva17, bonaduce20seastate}.

For users of the hindcast dataset, we encourage to subtract the hindcast mean storm surge water level for each grid point from the data before use, as we have done. Additionally, various methods for correction could be explored and applied to the hindcast dataset to further improve the quality. But this was not within the scope of the present work.

\section*{Declaration of competing interest}
The authors declare that they have no known competing financial interests or personal relationships that could have appeared to influence the work reported in this paper.

\section*{Acknowledgments}
{\O}B gratefully acknowledges the support by the Research Council of Norway through the Stormrisk project (grant no 300608). PT, JR, {\O}S and NMK gratefully acknowledge the support by the Research Council of Norway through the MachineOcean project (grant no 303411). The hindcast data are available at \url{https://thredds.met.no/thredds/catalog/stormrisk/catalog.html}. The NORA3 atmospheric forcing data are available at \url{https://thredds.met.no/thredds/projects/nora3.html}.
We would also like to thank Mari Hegland Halvorsen at the Norwegian Mapping Authorities for making the observations of water level for the year 1987 available to us.

\clearpage

\section*{Appendix A: Code and Data Release}

\subsection*{Observation data availability}

The license agreements of the water level observations we gathered do not allow to redistribute our dataset of in-situ observations. However, the dataset can be shared upon reasonable request and collaboration by taking contact with the corresponding author of this study. Moreover, the list of data sources presented in Table \ref{tab:obs_source} can be used to retrieve again the data. Though this involves a significant amount of work, we provide a series of code repositories that can be used as a basis or inspiration for building the aggregated dataset:

\begin{itemize}
    \item NO data: \url{https://github.com/jerabaul29/kartverket_storm_surge_data}
    \item WL data: \url{https://github.com/jerabaul29/d22_data_format_public}
    \item all other sources of data: \url{https://github.com/jerabaul29/external_water_level_stations}
    \item aggregation of the data into a single netcdf-cf dataset and computation of the modal tide analysis: \url{https://github.com/jerabaul29/storm_surge_aggregated_dataset}
    \item adapted pytide packet used in this study: \url{https://github.com/jerabaul29/pangeo-pytide/tree/add-example}
\end{itemize}

\section*{Appendix B: Observation data collection, quality control, and computation of the sea elevation residual}

\begin{table}[ht]
    \centering
    \begin{tabular}{|p{0.5in}|p{1.5in}|p{2.5in}|}
        \hline
         Prefix & Source & Data request URL \\
         \hline
         \hline
         NO & Kartverket - Norwegian Mapping Authority & \url{https://api.sehavniva.no/tideapi_en.html} \\
         \hline
         WL & Norwegian Meteorological Institute, oil platforms sensors & MetNo internal data: address data request to ocean@met.no \\
         \hline
         SW & Swedish Meteorological and Hydrological Institute & \url{https://www.smhi.se/data/oceanografi/ladda-ner-oceanografiska-observationer/#param=sealevelrh2000,stations=all} \\
         \hline
         DK & Danish Coastal Authority & \url{https://kyst.dk/hav-og-anlaeg/maalinger-og-data/vandstandsmaalinger} \\
         \hline
         DE & BFG - Federal Institute of Hydrology & \url{https://www.bafg.de/EN/06_Info_Service/01_WaterLevels/waterlevels_node.html} \\
         \hline
         NL & Rijkswaterstaat environment & \url{https://waterinfo.rws.nl/#/publiek/waterhoogte} \\
         \hline
         UK & BODC - British Oceanographic Data Center & \url{https://www.bodc.ac.uk/data/hosted_data_systems/sea_level/uk_tide_gauge_network/processed/} \\
         \hline
    \end{tabular}
    \caption{List of data sources for the water level observation data. The URLs are valid as of 2023-11, and may change in the future. The data format and access conditions depend from source to source.}
    \label{tab:obs_source}
\end{table}

\begin{figure}
    \centering
    \includegraphics[width=0.75\linewidth]{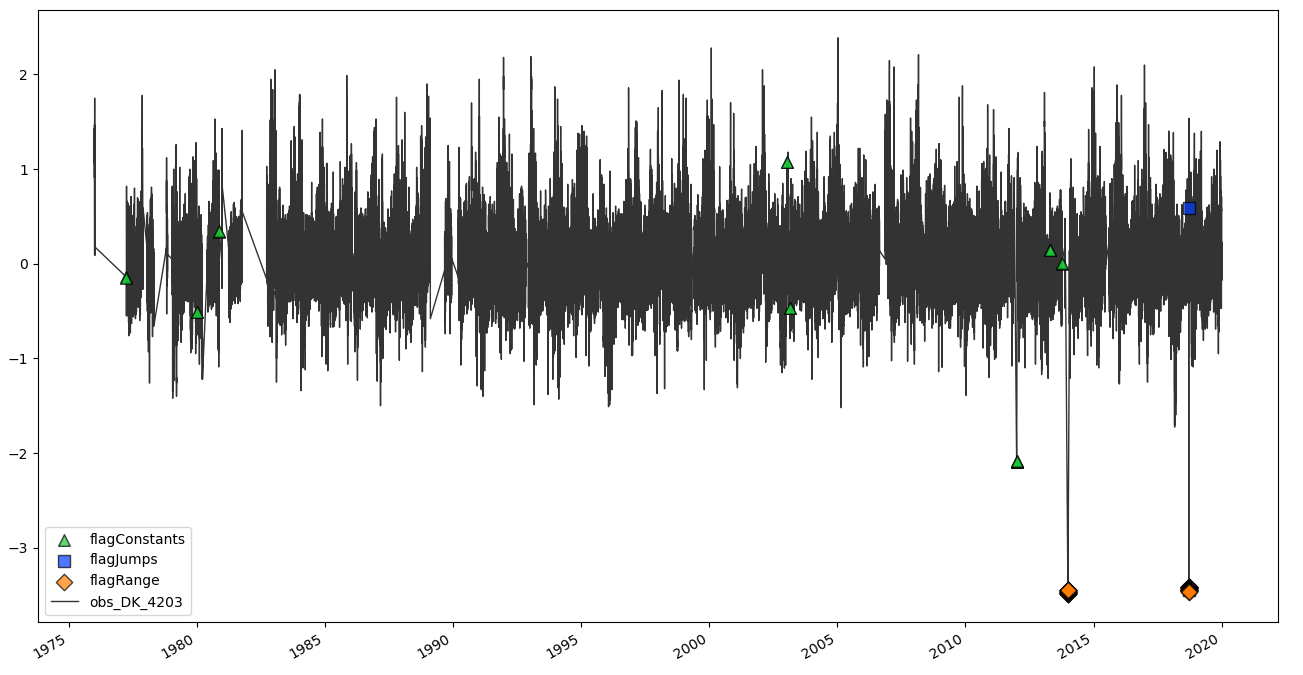}
    \caption{Time series of observed total water level from station DK\_4203 showing an example of different types of flagged values from the \acrshort{saqc} analysis.}
    \label{fig:saqc_example1}
\end{figure}

In order to assimilate the different observation data sources into a single homogeneous dataset, the following steps were taken:

\begin{itemize}
    \item For each data source, a custom parser (or even several parsers, as some of the data sources have format differences between different time periods and stations) was developed to convert the different data formats used into a list of tuples: $(time\_of\_observation, water\_level\_observation, quality\_flag)$.
    From there, a time series of the raw collected observation data is established for each station (note that the quality\_flag is available on a data source dependent basis).
    \item The time series for each station is, at this stage, a series of observation tuples, which temporal resolution varies between stations (and for some stations, with time), from 30 minutes (coarsest observations assimilated in our dataset) to 10 minutes (finest observations available for some stations). We perform a simple quality checking, and we first reject data that have either a faulty quality flag (when provided), or that have clearly faulty values (in particular unrealistic absolute value, such as for example values of 999.9 reported by some sensors as an error code; moreover, some instrument malfunctions can also result in readings that clearly stand out of the typical data range but with a less drastic absolute value, and these are detected later on by our quality control procedure, see illustration in Figure \ref{fig:saqc_example1}). The amount of such data points rejected varies from station to station, but typically less than 5\% of the data are flagged as such in the stations reported here. We then perform interpolation of these data onto a common time basis, with 10 minutes resolution (corresponding to UTC minutes equal to 0 modulo 10).
\end{itemize}

At this stage, consistent time series on a common time basis are obtained for all the stations. We then perform a second quality check, that is based on human investigation. In particular, the time series for each station was visually inspected for problems such as large average trends in the data (this happens, for example, for the oil platforms, that are standing on the seafloor and subject to seabed subsidence), and change of the reference levels following sensor replacement, maintenance, or re-positioning. These errors are detected by directly inspecting the signals, and the time series are then corrected into 0-mean data, either by linearly detrending the data between sensor re-positioning (at the oil station locations where subsidence takes place), or by computing the mean water elevation to subtract on segments of data in between sensor servicing when no subsidence is present.

Finally, we performed a third \acrfull{qc} of the data before using them to validate NORA-Surge, in order to ensure the datasets are as uniform as possible. This \acrshort{qc} is based on a second visual inspection and the open source software \acrshort{saqc} \citep{schmidtetal2023}. This has been applied to the DE, NL, DK, SW, and WL stations. We follow the steps described below:

\begin{enumerate}
    \item Visual inspection:
    \begin{enumerate}
        \item Discard stations with bad quality (e.g. very noisy or large amounts of missing data).
        \item Detect heteroscedasticity, and manually flag periods where the observations clearly have a different variance than the rest of the time series.
    \end{enumerate}
    \item Remove stations where the standard deviation of the model and observations differs by more than $10~cm$ (as shown in Figure \ref{fig:stddev_obs_mod}). This does not necessarily mean that the observations are erroneous, but is a strong indication that either the characteristics of the model does not resemble the observations, or vice versa, and hence we chose to discard these stations.
    \item At each location, given $\mu$ and $\sigma$ the mean and standard deviation of the observed values, respectively, we apply the following methods implemented in the \acrshort{saqc} library:
    \begin{enumerate}
        \item \texttt{flagUniLOF}, an implementation of the univariate Local Outlier Factor (LOF). The threshold is set to $\mathrm{thresh}=2.4 - \sigma$ and the number of periods to $n=6$.
        \item \texttt{flagJumps}, where the size of the two moving windows is $\mathrm{window}=3$ hours and the threshold value by which the mean of data has to jump is $\mathrm{thresh}=4\sigma$.
        \item \texttt{flagRange}, to flag values exceeding the closed interval $\mathrm{[min, max]}$, where $\mathrm{min}=\mu - 8\sigma$ and $\mathrm{max}=\mu + 8\sigma$.
        \item \texttt{flagConstants}, where the maximum total change allowed per window is set to $\mathrm{thresh}=0.01$ and the size of the moving window is $\mathrm{window}= 72$ hours.
        \item \texttt{flagByVariance} to flag low-variance data that do not exceed a threshold of $\mathrm{thresh}=\sigma/100$ in a rolling window with a size defined by $\mathrm{window}=72$.
        \item \texttt{flagIsolated} to flag temporal isolated groups of data. The gap window, this is, the minimum gap size required before and after a data group to consider it isolated, is set to $\mathrm{gap\_window}=3000$ hours, and a group window, this is, the maximum size of a data group to consider it a candidate for an isolated group, is set to $\mathrm{group\_window}=1000$ hours.
    \end{enumerate}
\end{enumerate}

\begin{figure}[ht]
    \centering
    \includegraphics[width=0.75\linewidth]{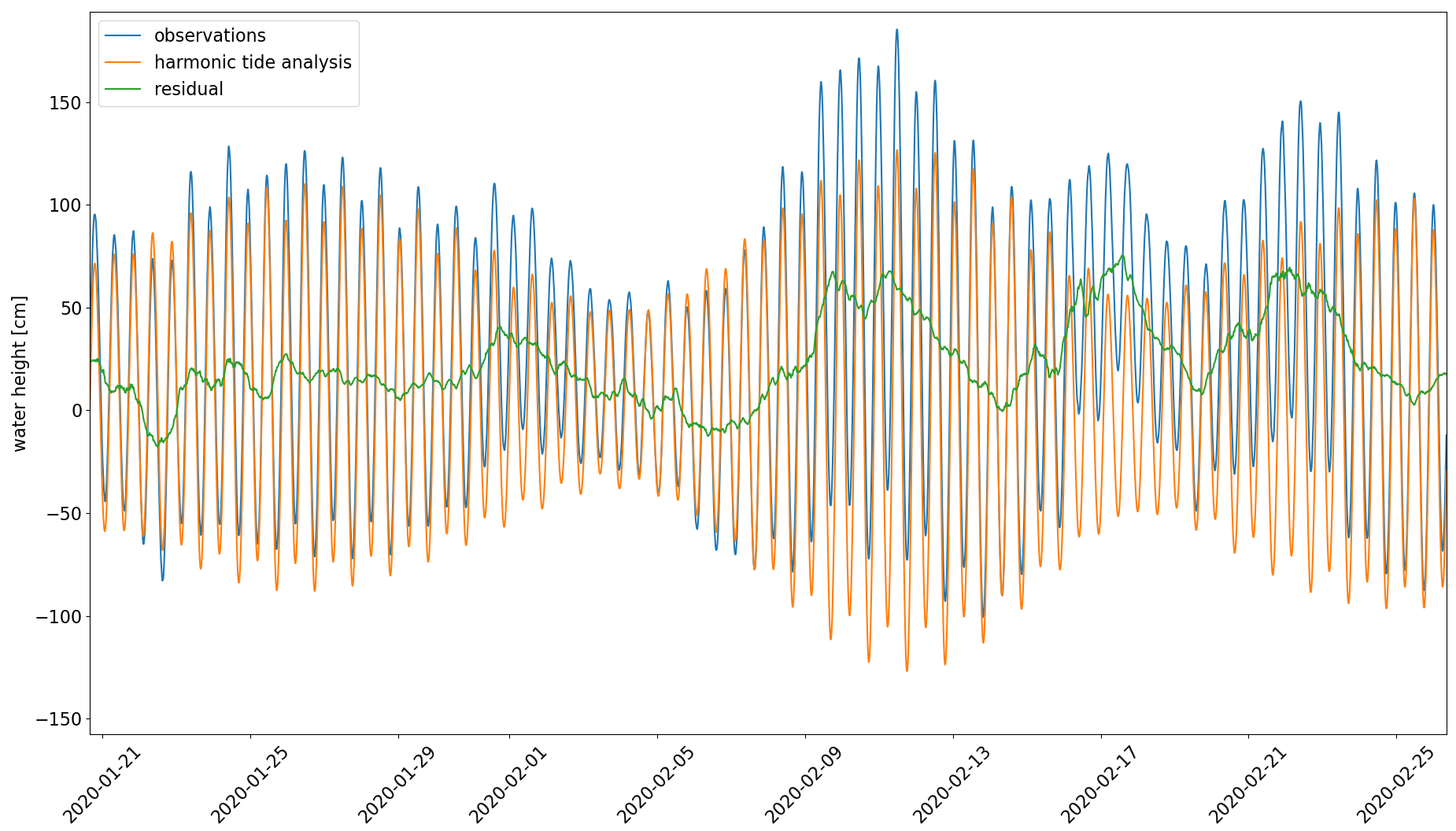}
    \caption{Illustration of the quality controlled observations, the harmonic tide analysis computed from the full duration of the observations timeseries using pytide, and the residual signal obtained by subtracting the former two, during part of the 2020 winter at the Mausund NO station. The figure is typically representative of what is obtained at other locations. The harmonic tide analysis captures well the tidal component of the sea level. The residual signal contains clear signatures of storms passing by the station, and the associated storm surges.}
    \label{fig:illustration_Mausund}
\end{figure}

At each station location, the full quality controlled sea elevation signal is then used to perform a standard harmonic tidal analysis using the pytide software tool, as described in the main body of the manuscript. This allows to obtain a best estimate for the weather-independent astronomical tide contribution to the sea level. Following this, we use the residual obtained by subtracting the harmonic tide contribution from the sea elevation observation to estimate the weather sea level surge component. This is illustrated, for a typically representative station (here, the Mausund NO station, though results obtained at other stations look similar), in Fig. \ref{fig:illustration_Mausund}. As visible there, the harmonic tide analysis captures the tide contribution well, and clear low-frequency weather effects corresponding to storms passing by and the associated storm surges are visible in the residual.

\printcredits

\bibliographystyle{cas-model2-names}

\bibliography{norass}

\begin{thebibliography}{33}
\expandafter\ifx\csname natexlab\endcsname\relax\def\natexlab#1{#1}\fi
\providecommand{\url}[1]{\texttt{#1}}
\providecommand{\href}[2]{#2}
\providecommand{\path}[1]{#1}
\providecommand{\DOIprefix}{doi:}
\providecommand{\ArXivprefix}{arXiv:}
\providecommand{\URLprefix}{URL: }
\providecommand{\Pubmedprefix}{pmid:}
\providecommand{\doi}[1]{\href{http://dx.doi.org/#1}{\path{#1}}}
\providecommand{\Pubmed}[1]{\href{pmid:#1}{\path{#1}}}
\providecommand{\bibinfo}[2]{#2}
\ifx\xfnm\relax \def\xfnm[#1]{\unskip,\space#1}\fi
\bibitem[{Andresen et~al.(1987)Andresen, Johnsen and
  Kristoffersen}]{andresen:etal:1987}
\bibinfo{author}{Andresen, L.}, \bibinfo{author}{Johnsen, {\O}.},
  \bibinfo{author}{Kristoffersen, D.}, \bibinfo{year}{1987}.
\newblock \bibinfo{title}{{U}været 16. og 17. oktober (forel{\o}pig rapport)}.
\newblock \bibinfo{type}{DNMI-Rapport} \bibinfo{number}{35/87}. Norwegian
  Meteorological Institute.
\bibitem[{Bernier et~al.(2024)Bernier, Hemer, Nobuhito, Appendini, Breivik, {de
  Camargo}, Casas-Prat, Duong, Haigh, Howard, Hernaman, Huizy, Irish, Kirezci,
  Kohno, Lee, McInnes, Meyer, Marcos, Marsooli, Oliva, Menendez, Moghimi, Muis,
  Polton, Pringle, Ranasinghe, Saillour, Smith, Tadesse, Swail, Tomoya,
  Voukouvalas, Wahl, Wang, Weisse, Westerink, Young and Zhang}]{bernier24}
\bibinfo{author}{Bernier, N.B.}, \bibinfo{author}{Hemer, M.},
  \bibinfo{author}{Nobuhito, M.}, \bibinfo{author}{Appendini, C.M.},
  \bibinfo{author}{Breivik, O.}, \bibinfo{author}{{de Camargo}, R.},
  \bibinfo{author}{Casas-Prat, M.}, \bibinfo{author}{Duong, T.M.},
  \bibinfo{author}{Haigh, I.D.}, \bibinfo{author}{Howard, T.},
  \bibinfo{author}{Hernaman, V.}, \bibinfo{author}{Huizy, O.},
  \bibinfo{author}{Irish, J.L.}, \bibinfo{author}{Kirezci, E.},
  \bibinfo{author}{Kohno, N.}, \bibinfo{author}{Lee, J.W.},
  \bibinfo{author}{McInnes, K.L.}, \bibinfo{author}{Meyer, E.},
  \bibinfo{author}{Marcos, M.}, \bibinfo{author}{Marsooli, R.},
  \bibinfo{author}{Oliva, A.M.}, \bibinfo{author}{Menendez, M.},
  \bibinfo{author}{Moghimi, S.}, \bibinfo{author}{Muis, S.},
  \bibinfo{author}{Polton, J.A.}, \bibinfo{author}{Pringle, W.J.},
  \bibinfo{author}{Ranasinghe, R.}, \bibinfo{author}{Saillour, T.},
  \bibinfo{author}{Smith, G.}, \bibinfo{author}{Tadesse, M.G.},
  \bibinfo{author}{Swail, V.}, \bibinfo{author}{Tomoya, S.},
  \bibinfo{author}{Voukouvalas, E.}, \bibinfo{author}{Wahl, T.},
  \bibinfo{author}{Wang, P.}, \bibinfo{author}{Weisse, R.},
  \bibinfo{author}{Westerink, J.J.}, \bibinfo{author}{Young, I.},
  \bibinfo{author}{Zhang, Y.J.}, \bibinfo{year}{2024}.
\newblock \bibinfo{title}{{Storm Surges and Extreme Sea Levels: Review,
  Establishment of Model Intercomparison and Coordination of Surge Climate
  Projection Efforts (SurgeMIP)}}.
\newblock \bibinfo{journal}{Weather and Climate Extremes} ,
  \bibinfo{pages}{100689}\DOIprefix\doi{10.1016/j.wace.2024.100689}.
\bibitem[{Bonaduce et~al.(2020)Bonaduce, Staneva, Grayek, Bidlot and
  Breivik}]{bonaduce20seastate}
\bibinfo{author}{Bonaduce, A.}, \bibinfo{author}{Staneva, J.},
  \bibinfo{author}{Grayek, S.}, \bibinfo{author}{Bidlot, J.R.},
  \bibinfo{author}{Breivik, {\O}.}, \bibinfo{year}{2020}.
\newblock \bibinfo{title}{{Sea-state contributions to sea-level variability in
  the European Seas}}.
\newblock \bibinfo{journal}{Ocean Dyn} \bibinfo{volume}{70},
  \bibinfo{pages}{1547--1569}.
\newblock \DOIprefix\doi{10.1007/s10236-020-01404-1}.
\bibitem[{Breivik et~al.(2022)Breivik, Carrasco, Haakenstad, Aarnes, Behrens,
  Bidlot, Bj{\"{o}}rkqvist, Bohlinger, Furevik, Staneva and Reistad}]{bre22}
\bibinfo{author}{Breivik, {\O}.}, \bibinfo{author}{Carrasco, A.},
  \bibinfo{author}{Haakenstad, H.}, \bibinfo{author}{Aarnes, O.J.},
  \bibinfo{author}{Behrens, A.}, \bibinfo{author}{Bidlot, J.R.},
  \bibinfo{author}{Bj{\"{o}}rkqvist, J.V.}, \bibinfo{author}{Bohlinger, P.},
  \bibinfo{author}{Furevik, B.R.}, \bibinfo{author}{Staneva, J.},
  \bibinfo{author}{Reistad, M.}, \bibinfo{year}{2022}.
\newblock \bibinfo{title}{{The Impact of a Reduced High-wind Charnock
  Coefficient on Wave Growth With Application to the North Sea, the Norwegian
  Sea and the Arctic Ocean}}.
\newblock \bibinfo{journal}{J Geophys Res: Oceans} \bibinfo{volume}{127},
  \bibinfo{pages}{e2021JC018196}.
\newblock \DOIprefix\doi{10.1029/2021JC018196}.
\bibitem[{Breivik et~al.(2015)Breivik, Mogensen, Bidlot, Balmaseda and
  Janssen}]{breivik2015}
\bibinfo{author}{Breivik, {\O}.}, \bibinfo{author}{Mogensen, K.},
  \bibinfo{author}{Bidlot, J.R.}, \bibinfo{author}{Balmaseda, M.A.},
  \bibinfo{author}{Janssen, P.A.E.M.}, \bibinfo{year}{2015}.
\newblock \bibinfo{title}{Surface wave effects in the nemo ocean model: Forced
  and coupled experiments}.
\newblock \bibinfo{journal}{Journal of Geophysical Research: Oceans}
  \bibinfo{volume}{120}, \bibinfo{pages}{2973--2992}.
\newblock \URLprefix
  \url{https://agupubs.onlinelibrary.wiley.com/doi/abs/10.1002/2014JC010565},
  \DOIprefix\doi{https://doi.org/10.1002/2014JC010565},
  \href{http://arxiv.org/abs/https://agupubs.onlinelibrary.wiley.com/doi/pdf/10.1002/2014JC010565}{\tt
  arXiv:https://agupubs.onlinelibrary.wiley.com/doi/pdf/10.1002/2014JC010565}.
\bibitem[{Chapman(1985)}]{chapman:1985}
\bibinfo{author}{Chapman, D.}, \bibinfo{year}{1985}.
\newblock \bibinfo{title}{Numerical treatment of cross-shelf open boundaries in
  a barotropic coastal ocean model}.
\newblock \bibinfo{journal}{J. Phys. Oceanogr.} \bibinfo{volume}{15},
  \bibinfo{pages}{1060--1075}.
\bibitem[{Charnock(1955)}]{charnock:1955}
\bibinfo{author}{Charnock, H.}, \bibinfo{year}{1955}.
\newblock \bibinfo{title}{Wind stress on a water surface}.
\newblock \bibinfo{journal}{Quarterly Journal of the Royal Meteorological
  Society} \bibinfo{volume}{81}, \bibinfo{pages}{639--640}.
\newblock \URLprefix
  \url{https://rmets.onlinelibrary.wiley.com/doi/abs/10.1002/qj.49708135027},
  \DOIprefix\doi{https://doi.org/10.1002/qj.49708135027},
  \href{http://arxiv.org/abs/https://rmets.onlinelibrary.wiley.com/doi/pdf/10.1002/qj.49708135027}{\tt
  arXiv:https://rmets.onlinelibrary.wiley.com/doi/pdf/10.1002/qj.49708135027}.
\bibitem[{Coles(2001)}]{col01}
\bibinfo{author}{Coles, S.}, \bibinfo{year}{2001}.
\newblock \bibinfo{title}{{An introduction to statistical modeling of extreme
  values}}.
\newblock \bibinfo{publisher}{Springer Verlag}.
\bibitem[{Deutschländer et~al.(2013)Deutschländer, Friedrich, Haeseler and
  Lefebvre}]{deutschlander:etal:2013}
\bibinfo{author}{Deutschländer, T.}, \bibinfo{author}{Friedrich, K.},
  \bibinfo{author}{Haeseler, S.}, \bibinfo{author}{Lefebvre, C.},
  \bibinfo{year}{2013}.
\newblock \bibinfo{title}{Severe storm XAVER across northern Europe from 5 to 7
  December 2013}.
\newblock \bibinfo{type}{DWD Report}. Deutscher Wetterdienst.
\bibitem[{Engen(1988)}]{engen:1988}
\bibinfo{author}{Engen, I.K.}, \bibinfo{year}{1988}.
\newblock \bibinfo{title}{{F}lommen på {S}{\o}r- og {{\O}}stlandet i oktober
  1987}.
\newblock \bibinfo{type}{NVE Publikasjon} \bibinfo{number}{15}. Norwegian Water
  Resources and Energy Directorate.
\bibitem[{Fernández-Montblanc et~al.(2020)Fernández-Montblanc, Vousdoukas,
  Mentaschi and Ciavola}]{FERNANDEZMONTBLANC2020105367}
\bibinfo{author}{Fernández-Montblanc, T.}, \bibinfo{author}{Vousdoukas, M.},
  \bibinfo{author}{Mentaschi, L.}, \bibinfo{author}{Ciavola, P.},
  \bibinfo{year}{2020}.
\newblock \bibinfo{title}{A pan-european high resolution storm surge hindcast}.
\newblock \bibinfo{journal}{Environment International} \bibinfo{volume}{135},
  \bibinfo{pages}{105367}.
\newblock \URLprefix
  \url{https://www.sciencedirect.com/science/article/pii/S0160412019324055},
  \DOIprefix\doi{https://doi.org/10.1016/j.envint.2019.105367}.
\bibitem[{Flather(1976)}]{flather:1976}
\bibinfo{author}{Flather, R.A.}, \bibinfo{year}{1976}.
\newblock \bibinfo{title}{A tidal model of the northwest {E}uropean continental
  shelf.}
\newblock \bibinfo{journal}{Memoires de la Societe Royale de Sciences de Liege}
  \bibinfo{volume}{6}, \bibinfo{pages}{141--164}.
\bibitem[{Glahn et~al.(2009)Glahn, Taylor, Kurkowski and
  Shaffer}]{glahn2009role}
\bibinfo{author}{Glahn, B.}, \bibinfo{author}{Taylor, A.},
  \bibinfo{author}{Kurkowski, N.}, \bibinfo{author}{Shaffer, W.A.},
  \bibinfo{year}{2009}.
\newblock \bibinfo{title}{The role of the slosh model in national weather
  service storm surge forecasting}.
\newblock \bibinfo{journal}{National Weather Digest} \bibinfo{volume}{33},
  \bibinfo{pages}{3--14}.
\bibitem[{Gumbel(1958)}]{gumbel1958statistics}
\bibinfo{author}{Gumbel, E.J.}, \bibinfo{year}{1958}.
\newblock \bibinfo{title}{Statistics of extremes}.
\newblock \bibinfo{publisher}{Columbia university press}.
\bibitem[{Haakenstad and Breivik(2022)}]{haakenstad22}
\bibinfo{author}{Haakenstad, H.}, \bibinfo{author}{Breivik, {\O}.},
  \bibinfo{year}{2022}.
\newblock \bibinfo{title}{{NORA3 Part II: Precipitation and temperature
  statistics in complex terrain modeled with a non-hydrostatic model}}.
\newblock \bibinfo{journal}{J Appl Meteor and Climatol} \bibinfo{volume}{61},
  \bibinfo{pages}{1549--1572}.
\newblock \DOIprefix\doi{10.1175/JAMC-D-22-0005.1}.
\bibitem[{Haakenstad et~al.(2021)Haakenstad, Breivik, Furevik, Reistad,
  Bohlinger and Aarnes}]{haakenstad21nora3}
\bibinfo{author}{Haakenstad, H.}, \bibinfo{author}{Breivik, {\O}.},
  \bibinfo{author}{Furevik, B.}, \bibinfo{author}{Reistad, M.},
  \bibinfo{author}{Bohlinger, P.}, \bibinfo{author}{Aarnes, O.J.},
  \bibinfo{year}{2021}.
\newblock \bibinfo{title}{{NORA3: A nonhydrostatic high-resolution hindcast of
  the North Sea, the Norwegian Sea, and the Barents Sea}}.
\newblock \bibinfo{journal}{J Appl Meteor and Climatol} \bibinfo{volume}{60},
  \bibinfo{pages}{1443--1464}.
\newblock \DOIprefix\doi{10.1175/JAMC-D-21-0029.1}.
\bibitem[{Haidvogel et~al.(2008)Haidvogel, Arango, Budgell, Cornuelle,
  Curchitser, Lorenzo, Fennel, Geyer, Hermann, Lanerolle, Levin, McWilliams,
  Miller, Moore, Powell, Shchepetkin, Sherwood, Signell, Warner and
  Wilkin}]{haidv:etal:2008}
\bibinfo{author}{Haidvogel, D.B.}, \bibinfo{author}{Arango, H.},
  \bibinfo{author}{Budgell, P.W.}, \bibinfo{author}{Cornuelle, B.D.},
  \bibinfo{author}{Curchitser, E.}, \bibinfo{author}{Lorenzo, E.D.},
  \bibinfo{author}{Fennel, K.}, \bibinfo{author}{Geyer, W.R.},
  \bibinfo{author}{Hermann, A.J.}, \bibinfo{author}{Lanerolle, L.},
  \bibinfo{author}{Levin, J.}, \bibinfo{author}{McWilliams, J.C.},
  \bibinfo{author}{Miller, A.J.}, \bibinfo{author}{Moore, A.M.},
  \bibinfo{author}{Powell, T.M.}, \bibinfo{author}{Shchepetkin, A.F.},
  \bibinfo{author}{Sherwood, C.R.}, \bibinfo{author}{Signell, R.P.},
  \bibinfo{author}{Warner, J.C.}, \bibinfo{author}{Wilkin, J.},
  \bibinfo{year}{2008}.
\newblock \bibinfo{title}{Ocean forecasting in terrain-following coordinates:
  Formulation and skill assessment of the {R}egional {O}cean {M}odeling
  {S}ystem}.
\newblock \bibinfo{journal}{J. Comput. Phys.} \bibinfo{volume}{227},
  \bibinfo{pages}{3595--3624}.
\newblock \DOIprefix\doi{http://dx.doi.org/10.1016/j.jcp.2007.06.016}.
\bibitem[{Hersbach et~al.(2020)Hersbach, Bell, Berrisford, Hirahara, Horányi,
  Muñoz-Sabater, Nicolas, Peubey, Radu, Schepers, Simmons, Soci, Abdalla,
  Abellan, Balsamo, Bechtold, Biavati, Bidlot, Bonavita, De~Chiara, Dahlgren,
  Dee, Diamantakis, Dragani, Flemming, Forbes, Fuentes, Geer, Haimberger,
  Healy, Hogan, Hólm, Janisková, Keeley, Laloyaux, Lopez, Lupu, Radnoti,
  de~Rosnay, Rozum, Vamborg, Villaume and Thépaut}]{hersbach20era5}
\bibinfo{author}{Hersbach, H.}, \bibinfo{author}{Bell, B.},
  \bibinfo{author}{Berrisford, P.}, \bibinfo{author}{Hirahara, S.},
  \bibinfo{author}{Horányi, A.}, \bibinfo{author}{Muñoz-Sabater, J.},
  \bibinfo{author}{Nicolas, J.}, \bibinfo{author}{Peubey, C.},
  \bibinfo{author}{Radu, R.}, \bibinfo{author}{Schepers, D.},
  \bibinfo{author}{Simmons, A.}, \bibinfo{author}{Soci, C.},
  \bibinfo{author}{Abdalla, S.}, \bibinfo{author}{Abellan, X.},
  \bibinfo{author}{Balsamo, G.}, \bibinfo{author}{Bechtold, P.},
  \bibinfo{author}{Biavati, G.}, \bibinfo{author}{Bidlot, J.},
  \bibinfo{author}{Bonavita, M.}, \bibinfo{author}{De~Chiara, G.},
  \bibinfo{author}{Dahlgren, P.}, \bibinfo{author}{Dee, D.},
  \bibinfo{author}{Diamantakis, M.}, \bibinfo{author}{Dragani, R.},
  \bibinfo{author}{Flemming, J.}, \bibinfo{author}{Forbes, R.},
  \bibinfo{author}{Fuentes, M.}, \bibinfo{author}{Geer, A.},
  \bibinfo{author}{Haimberger, L.}, \bibinfo{author}{Healy, S.},
  \bibinfo{author}{Hogan, R.J.}, \bibinfo{author}{Hólm, E.},
  \bibinfo{author}{Janisková, M.}, \bibinfo{author}{Keeley, S.},
  \bibinfo{author}{Laloyaux, P.}, \bibinfo{author}{Lopez, P.},
  \bibinfo{author}{Lupu, C.}, \bibinfo{author}{Radnoti, G.},
  \bibinfo{author}{de~Rosnay, P.}, \bibinfo{author}{Rozum, I.},
  \bibinfo{author}{Vamborg, F.}, \bibinfo{author}{Villaume, S.},
  \bibinfo{author}{Thépaut, J.N.}, \bibinfo{year}{2020}.
\newblock \bibinfo{title}{The era5 global reanalysis}.
\newblock \bibinfo{journal}{Quarterly Journal of the Royal Meteorological
  Society} \bibinfo{volume}{146}, \bibinfo{pages}{1999--2049}.
\newblock \URLprefix
  \url{https://rmets.onlinelibrary.wiley.com/doi/abs/10.1002/qj.3803},
  \DOIprefix\doi{https://doi.org/10.1002/qj.3803},
  \href{http://arxiv.org/abs/https://rmets.onlinelibrary.wiley.com/doi/pdf/10.1002/qj.3803}{\tt
  arXiv:https://rmets.onlinelibrary.wiley.com/doi/pdf/10.1002/qj.3803}.
\bibitem[{Horsburgh and Wilson(2007)}]{horsburgh:2007}
\bibinfo{author}{Horsburgh, K.J.}, \bibinfo{author}{Wilson, C.},
  \bibinfo{year}{2007}.
\newblock \bibinfo{title}{Tide-surge interaction and its role in the
  distribution of surge residuals in the north sea}.
\newblock \bibinfo{journal}{Journal of Geophysical Research: Oceans}
  \bibinfo{volume}{112}.
\newblock \URLprefix
  \url{https://agupubs.onlinelibrary.wiley.com/doi/abs/10.1029/2006JC004033},
  \DOIprefix\doi{https://doi.org/10.1029/2006JC004033},
  \href{http://arxiv.org/abs/https://agupubs.onlinelibrary.wiley.com/doi/pdf/10.1029/2006JC004033}{\tt
  arXiv:https://agupubs.onlinelibrary.wiley.com/doi/pdf/10.1029/2006JC004033}.
\bibitem[{Idier et~al.(2019)Idier, Bertin, Thompson and Pickering}]{idier:2019}
\bibinfo{author}{Idier, D.}, \bibinfo{author}{Bertin, X.},
  \bibinfo{author}{Thompson, P.}, \bibinfo{author}{Pickering, M.D.},
  \bibinfo{year}{2019}.
\newblock \bibinfo{title}{Interactions between mean sea level, tide, surge,
  waves and flooding: Mechanisms and contributions to sea level variations at
  the coast}.
\newblock \bibinfo{journal}{Surveys in Geophysics} \bibinfo{volume}{40},
  \bibinfo{pages}{1603--1630}.
\newblock \DOIprefix\doi{https://doi.org/10.1007/s10712-019-09549-5}.
\bibitem[{Kierulf et~al.(2021)Kierulf, van Pelt, Petrov, Dähnn, Kirkvik and
  Omang}]{kierulf2021}
\bibinfo{author}{Kierulf, H.P.}, \bibinfo{author}{van Pelt, W.J.J.},
  \bibinfo{author}{Petrov, L.}, \bibinfo{author}{Dähnn, M.},
  \bibinfo{author}{Kirkvik, A.S.}, \bibinfo{author}{Omang, O.},
  \bibinfo{year}{2021}.
\newblock \bibinfo{title}{{Seasonal glacier and snow loading in Svalbard
  recovered from geodetic observations}}.
\newblock \bibinfo{journal}{Geophysical Journal International}
  \bibinfo{volume}{229}, \bibinfo{pages}{408--425}.
\newblock \URLprefix \url{https://doi.org/10.1093/gji/ggab482},
  \DOIprefix\doi{10.1093/gji/ggab482},
  \href{http://arxiv.org/abs/https://academic.oup.com/gji/article-pdf/229/1/408/42020938/ggab482.pdf}{\tt
  arXiv:https://academic.oup.com/gji/article-pdf/229/1/408/42020938/ggab482.pdf}.
\bibitem[{Kristensen et~al.(2022)Kristensen, R{\o}ed and
  S{\ae}tra}]{kristensen22}
\bibinfo{author}{Kristensen, N.}, \bibinfo{author}{R{\o}ed, L.},
  \bibinfo{author}{S{\ae}tra, {\O}.}, \bibinfo{year}{2022}.
\newblock \bibinfo{title}{{A forecasting and warning system of storm surge
  events along the Norwegian coast}}.
\newblock \bibinfo{journal}{Environ Fluid Mech}
  \DOIprefix\doi{10.1007/s10652-022-09871-4}.
\bibitem[{Melet et~al.(2018)Melet, Meyssignac, Almar and Le~Cozannet}]{melet18}
\bibinfo{author}{Melet, A.}, \bibinfo{author}{Meyssignac, B.},
  \bibinfo{author}{Almar, R.}, \bibinfo{author}{Le~Cozannet, G.},
  \bibinfo{year}{2018}.
\newblock \bibinfo{title}{{Under-estimated wave contribution to coastal
  sea-level rise}}.
\newblock \bibinfo{journal}{Nature Climate Change} \bibinfo{volume}{8},
  \bibinfo{pages}{234--239}.
\newblock \DOIprefix\doi{10.1038/s41558-018-0088-y}.
\bibitem[{Saetra et~al.(2007)Saetra, Albretsen and Janssen}]{saetra2007}
\bibinfo{author}{Saetra, {\O}.}, \bibinfo{author}{Albretsen, J.},
  \bibinfo{author}{Janssen, P.A.E.M.}, \bibinfo{year}{2007}.
\newblock \bibinfo{title}{Sea-state-dependent momentum fluxes for ocean
  modeling}.
\newblock \bibinfo{journal}{Journal of Physical Oceanography}
  \bibinfo{volume}{37}, \bibinfo{pages}{2714 -- 2725}.
\newblock \URLprefix
  \url{https://journals.ametsoc.org/view/journals/phoc/37/11/2007jpo3582.1.xml},
  \DOIprefix\doi{10.1175/2007JPO3582.1}.
\bibitem[{Schmidt et~al.(2023)Schmidt, Sch{\"a}fer, Geller, L{\"u}nenschloss,
  Palm, Rinke, Rebmann, Rode and Bumberger}]{schmidtetal2023}
\bibinfo{author}{Schmidt, L.}, \bibinfo{author}{Sch{\"a}fer, D.},
  \bibinfo{author}{Geller, J.}, \bibinfo{author}{L{\"u}nenschloss, P.},
  \bibinfo{author}{Palm, B.}, \bibinfo{author}{Rinke, K.},
  \bibinfo{author}{Rebmann, C.}, \bibinfo{author}{Rode, M.},
  \bibinfo{author}{Bumberger, J.}, \bibinfo{year}{2023}.
\newblock \bibinfo{title}{System for automated quality control (saqc) to enable
  traceable and reproducible data streams in environmental science}.
\newblock \bibinfo{journal}{Environmental Modelling \& Software}
  \bibinfo{volume}{169}, \bibinfo{pages}{105809}.
\bibitem[{Selberg et~al.(2020)Selberg, Skjerdal, Julsrud, {\O}iestad, Andersen
  and Kristensen}]{selberg:etal:2020}
\bibinfo{author}{Selberg, L.}, \bibinfo{author}{Skjerdal, M.},
  \bibinfo{author}{Julsrud, I.}, \bibinfo{author}{{\O}iestad, M.},
  \bibinfo{author}{Andersen, A.}, \bibinfo{author}{Kristensen, N.},
  \bibinfo{year}{2020}.
\newblock \bibinfo{title}{{MET} {I}nfo {H}endelsesrapport, {E}kstremv{\ae}ret
  {E}lsa mandag 10. og tirsdag 11. februar 2020}.
\newblock \bibinfo{type}{MET Info} \bibinfo{number}{17}. Norwegian
  Meteorological Institute.
\bibitem[{Shchepetkin and McWilliams(2005)}]{shchepetkin05}
\bibinfo{author}{Shchepetkin, A.F.}, \bibinfo{author}{McWilliams, J.C.},
  \bibinfo{year}{2005}.
\newblock \bibinfo{title}{{The regional oceanic modeling system (ROMS): a
  split-explicit, free-surface, topography-following-coordinate oceanic
  model}}.
\newblock \bibinfo{journal}{Ocean Model} \bibinfo{volume}{9},
  \bibinfo{pages}{347--404}.
\newblock \DOIprefix\doi{10.1016/j.ocemod.2004.08.002}.
\bibitem[{Skjerdal et~al.(2020)Skjerdal, Olsen, Andersen and
  Kristensen}]{skjerdal:etal:2020}
\bibinfo{author}{Skjerdal, M.}, \bibinfo{author}{Olsen, A.M.},
  \bibinfo{author}{Andersen, A.}, \bibinfo{author}{Kristensen, N.},
  \bibinfo{year}{2020}.
\newblock \bibinfo{title}{{MET} {I}nfo {H}endelsesrapport, {E}kstremv{\ae}ret
  {D}idrik onsdag 15. januar 2020}.
\newblock \bibinfo{type}{MET Info} \bibinfo{number}{15}. Norwegian
  Meteorological Institute.
\bibitem[{Staneva et~al.(2017)Staneva, Alari, Breivik, Bidlot and
  Mogensen}]{staneva17}
\bibinfo{author}{Staneva, J.}, \bibinfo{author}{Alari, V.},
  \bibinfo{author}{Breivik, {\O}.}, \bibinfo{author}{Bidlot, J.R.},
  \bibinfo{author}{Mogensen, K.}, \bibinfo{year}{2017}.
\newblock \bibinfo{title}{{Effects of wave-induced forcing on a circulation
  model of the North Sea}}.
\newblock \bibinfo{journal}{Ocean Dyn} \bibinfo{volume}{67},
  \bibinfo{pages}{81--101}.
\newblock \DOIprefix\doi{10.1007/s10236-016-1009-0}. \bibinfo{note}{14th wave
  special issue}.
\bibitem[{Stockdon et~al.(2006)Stockdon, Holman, Howd and
  Sallenger}]{stockdon06}
\bibinfo{author}{Stockdon, H.}, \bibinfo{author}{Holman, R.},
  \bibinfo{author}{Howd, P.}, \bibinfo{author}{Sallenger, A.},
  \bibinfo{year}{2006}.
\newblock \bibinfo{title}{{Empirical parameterization of setup, swash, and
  runup}}.
\newblock \bibinfo{journal}{Coast Eng} \bibinfo{volume}{53},
  \bibinfo{pages}{573--588}.
\newblock \DOIprefix\doi{10.1016/j.coastaleng.2005.12.005}.
\bibitem[{Tedesco et~al.(2024)Tedesco, Rabault, Sætra, Kristensen, Aarnes,
  Breivik, Mauritzen and Sætra}]{tedesco23}
\bibinfo{author}{Tedesco, P.}, \bibinfo{author}{Rabault, J.},
  \bibinfo{author}{Sætra, M.L.}, \bibinfo{author}{Kristensen, N.M.},
  \bibinfo{author}{Aarnes, O.J.}, \bibinfo{author}{Breivik, {\O}.},
  \bibinfo{author}{Mauritzen, C.}, \bibinfo{author}{Sætra, {\O}.},
  \bibinfo{year}{2024}.
\newblock \bibinfo{title}{Bias correction of operational storm surge forecasts
  using neural networks}.
\newblock \bibinfo{journal}{Ocean Modelling} \bibinfo{volume}{188},
  \bibinfo{pages}{102334}.
\newblock \URLprefix
  \url{https://www.sciencedirect.com/science/article/pii/S1463500324000210},
  \DOIprefix\doi{https://doi.org/10.1016/j.ocemod.2024.102334}.
\bibitem[{Wang and Bernier(2023)}]{wang2023}
\bibinfo{author}{Wang, P.}, \bibinfo{author}{Bernier, N.B.},
  \bibinfo{year}{2023}.
\newblock \bibinfo{title}{Adding sea ice effects to a global operational model
  (nemo v3.6) for forecasting total water level: approach and impact}.
\newblock \bibinfo{journal}{Geoscientific Model Development}
  \bibinfo{volume}{16}, \bibinfo{pages}{3335--3354}.
\newblock \URLprefix \url{https://gmd.copernicus.org/articles/16/3335/2023/},
  \DOIprefix\doi{10.5194/gmd-16-3335-2023}.
\bibitem[{Zijl et~al.(2013)Zijl, Verlaan and Gerritsen}]{zijl:etal:2013}
\bibinfo{author}{Zijl, F.}, \bibinfo{author}{Verlaan, M.},
  \bibinfo{author}{Gerritsen, H.}, \bibinfo{year}{2013}.
\newblock \bibinfo{title}{Improved water-level forecasting for the {N}orthwest
  {E}uropean {S}helf and {N}orth {S}ea through direct modelling of tide, surge
  and non-linear interaction.}
\newblock \bibinfo{journal}{Ocean Dynamics} \bibinfo{volume}{63},
  \bibinfo{pages}{823–847}.
\newblock \DOIprefix\doi{https://doi.org/10.1007/s10236-013-0624-2}.

\end{thebibliography}

\end{document}